\documentclass[twocolumn,prl,10pt,aps,floatfix]{revtex4-1}
\usepackage[english]{babel}
\usepackage{graphicx}
\usepackage{amsmath}
\usepackage[colorlinks,linkcolor=blue,citecolor=blue,urlcolor=blue]{hyperref}
\usepackage{color}
\usepackage{soul}
\definecolor{nblue} {RGB}{28,130,185}
\let\oldaddcontentsline\addcontentsline
\newcommand{\stoptocentries}{\renewcommand{\addcontentsline}[3]{}}
\newcommand{\starttocentries}{\let\addcontentsline\oldaddcontentsline}
\begin{document}
\title{Interaction-induced topological phase transition and Majorana edge states\\in low-dimensional orbital-selective Mott insulators}
\author{J. Herbrych$^{1}$}
\author{M. {\'S}roda$^{1}$}
\author{G. Alvarez$^{2}$}
\author{M. Mierzejewski$^{1}$}
\author{E. Dagotto$^{3,4}$}
\affiliation{$^{1}$Department of Theoretical Physics, Faculty of Fundamental Problems of Technology, Wroc{\l}aw University of Science and Technology, 50-370 Wroc{\l}aw, Poland}
\affiliation{$^{2}$Computational Sciences and Engineering Division and Center for Nanophase Materials Sciences, Oak Ridge National Laboratory, Oak Ridge, Tennessee 37831, USA}
\affiliation{$^{3}$Department of Physics and Astronomy, University of Tennessee, Knoxville, Tennessee 37996, USA}
\affiliation{$^{4}$Materials Science and Technology Division, Oak Ridge National Laboratory, Oak Ridge, Tennessee 37831, USA}
\date{\today}
\begin{abstract}
Topological phases of matter are among the most intriguing research directions in Condensed Matter Physics. It is known that superconductivity induced on a topological insulator's surface can lead to exotic Majorana modes, the main ingredient of many proposed quantum computation schemes. In this context, the iron-based high critical temperature superconductors are a promising platform to host such an exotic phenomenon in real condensed-matter compounds. The Coulomb interaction is commonly believed to be vital for the magnetic and superconducting properties of these systems. This work bridges these two perspectives and shows that the Coulomb interaction can also drive a canonical superconductor with orbital degrees of freedom into the topological state. Namely, we show that above a critical value of the Hubbard interaction the system simultaneously develops spiral spin order, a highly unusual triplet amplitude in superconductivity, and, remarkably, Majorana fermions at the edges of the system.
\end{abstract}
\maketitle
\stoptocentries
%================================================================================

Topologically protected Majorana fermions - the elusive particles which are their own antiparticles - are exciting because of their potential importance in fault-resistant quantum computation. From the experimental perspective, heterostructure-based setups were proposed as the main candidates to host the Majorana zero-energy modes (MZM). For example, the topologically protected gapless surface states of topological insulators can be promoted to MZM by the proximity-induced pairing caused by an underlying superconducting (SC) substrate~\cite{Elliott2015}. However, the large spin-orbit coupling required to split the doubly-degenerated bands due to the electronic spins, renders such a setup hard to engineer. Another group of proposals utilizes magnetic atoms (e.g., Gd, Cr, or Fe) arranged in a chain structure on a BCS superconductor \cite{Braunecker2010,Perge2013,Braunecker2013,Klinovaja2013,Vazifeh2013,nadjperg2014,Steinbrecher2018,Kim2018,Jacek2019,Palacio2019,Pawlak2019}. These important efforts have shown that creating MZM in real condensed-matter compounds is challenging and only rare examples are currently available.

\begin{figure*}[!htb]
\includegraphics[width=1.0\textwidth]{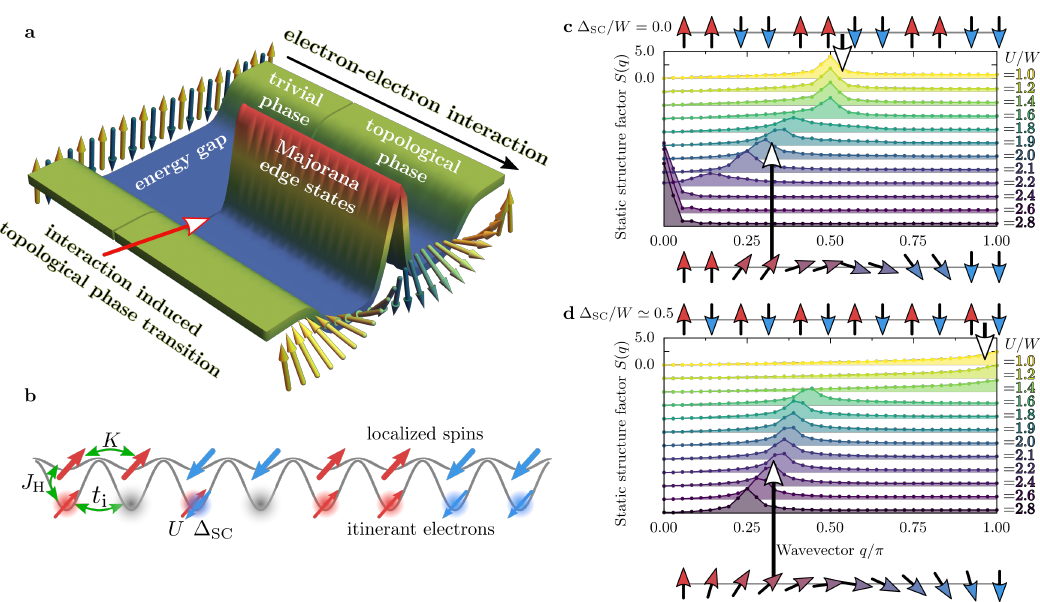}
\caption{{\bf Orbital-selective Mott phase and Majoranas.} {\bf a} Sketch of the chain edge density-of-states as function of the electron-electron Hubbard interaction strength. Magnetic orders (depicted by arrows) in the trivial and the topological superconducting (SC) phases are also presented. {\bf b} Schematic representation of the generalized Kondo-Heisenberg model studied here, with localized and itinerant electrons and simultaneously active Hubbard $U$ and superconducting $\Delta_{\mathrm{SC}}$ couplings. {\bf c-d} Interaction $U$ dependence of the static structure factor $S(q)$ for {\bf c} $\Delta_{\mathrm{SC}}=0$, {\bf d} $\Delta_{\mathrm{SC}}/W\simeq0.5$ (calculated for $L=36$ and $\overline{n}=0.5$).}
\label{fig1}
\end{figure*}

Interestingly, a series of recent works have shown that doped high critical temperature iron-based superconductor Fe(Se,Te) can host MZM~\cite{Wang2018,Zhang2018,Machida2019,Wang2020,Chen2020} . Although the electron-electron interaction is believed to be relevant for the pairing, its role in the stabilization of MZM is unknown. In fact, in most theoretical proposals to realize MZM, these zero-energy modes are a consequence of specific features in the non-interacting band structure, with the electron-electron interaction playing only a secondary role (and often even destabilizing the MZM)~\cite{Thomale2013,Rahmani2015}. By contrast, here we will show that a superconducting system with orbital degrees of freedom can be driven into a topologically nontrivial phase hosting MZM via increasing Hubbard interactions; see illustrative sketch in Fig.~\ref{fig1}a. We will focus on a generic model with coexisting wide and narrow energy bands, relevant to low-dimensional iron-based materials~\cite{Daghofer2010}. It was previously shown \cite{Rincon2014-1,Patel2016,Herbrych2018} that the multi-orbital Hubbard model can accurately capture static and dynamical properties of iron selenides, especially the block-magnetic order~\cite{Mourigal2015} of the 123 family AFe$_2$X$_3$ of iron-based ladders (with A alkali metals and X chalcogenides). For example, the three- and two-orbital Hubbard model on a one-dimensional (1D) lattice~\cite{Herbrych2018,Herbrych2020} successfully reproduces the inelastic neutron scattering spin spectrum, with nontrivial optical and acoustic modes. The aforementioned models exhibit~\cite{Rincon2014-1,Herbrych2019-1} the orbital-selective Mott phase (OSMP), with coexistent Mott-localized electrons in one orbital and itinerant electrons in the remaining orbitals. The system is then in an exotic state with simultaneously metallic and insulating properties. Furthermore, the localized orbitals have vanishing charge fluctuations, simplifying the description~\cite{Herbrych2019-1} into an OSMP effective model, i.e. the generalized Kondo-Heisenberg model (gKH)
\begin{eqnarray}
H_{\mathrm{gKH}}&=&t_{\mathrm{i}}\sum_{\ell,\sigma}
\left(c^{\dagger}_{\ell,\sigma}c^{\phantom{\dagger}}_{\ell+1,\sigma}+\mathrm{H.c.}\right)
+U\sum_{\ell}n_{\ell,\uparrow}n_{\ell,\downarrow}\nonumber\\
&+&\mu\sum_{\ell,\sigma}n_{\ell,\sigma}-2J_{\mathrm{H}}\sum_{\ell}\mathbf{S}_{\ell} \cdot \mathbf{s}_{\ell}
+K\sum_{\ell}\mathbf{S}_{\ell} \cdot \mathbf{S}_{\ell+1}\,.
\label{hamkon}
\end{eqnarray}
The first three terms in the above Hamiltonian describe the itinerant electrons: $c^{\dagger}_{\ell,\sigma}$ ($c^{\phantom{\dagger}}_{\ell,\sigma}$) creates (destroys) an electron with spin projection $\sigma=\{\uparrow,\downarrow\}$ at site $\ell=\{1,\dots,L\}$, $t_{\mathrm{i}}$ is their hopping amplitude, $U$ is the repulsive Hubbard interaction, and $\mu=\epsilon_F$ is the Fermi energy set by the density of itinerant electrons $\overline{n}=\sum_\ell(n_{\ell,\uparrow}+n_{\ell,\downarrow})/L$.

%--------------------------------------------------------------------------------
%--------------------------------------------------------------------------------

The double occupancy of the localized orbital can be eliminated by the Schrieffer-Wolff transformation and the remaining degrees of freedom, the localized spins $\mathbf{S}_{\ell}$ in the above model, interact with one another via a Heisenberg term with spin exchange $K=4t_{\mathrm{l}}^2/U$ [$t_{\mathrm{l}}$ is the hopping amplitude within the localized band]. Finally, $J_\mathrm{H}$ stands for the on-site interorbital Hund interaction, coupling the spins of the localized and itinerant electrons, $\mathbf{S}_{\ell}$ and $\mathbf{s}_{\ell}$, respectively. Figure~\ref{fig1}b contains a sketch of the model. Here, we consider a 1D lattice and use $t_{\mathrm{i}}=0.5$~[eV] and $t_{\mathrm{l}}=0.15$~[eV], with kinetic energy bandwidth $W=2.1$~[eV] as unit of energy~\cite{Herbrych2019-2}. Furthermore, to reduce the number of parameters in the model, we set $J_{\mathrm{H}}/U=1/4$, value widely used when modeling iron superconductors. Systems with open boundary conditions are studied via the density-matrix renormalization group (DMRG) method (see Methods section).

The key ingredient in systems expected to host the MZM~\cite{Kitaev2001} is the presence of a SC gap, modeled typically by an $s$-wave pairing field. Such a term represents the proximity effect \cite{vanWees1997} induced on the magnetic system by an external $s$-wave superconductor. However, it should be noted that the SC proximity effect has to be considered with utmost care. For example, recent experimental investigations~\cite{Hlevyack2020} showed that although the interface between Nb (BCS $s$-wave SC) and Bi$_2$Se$_3$ film (topological metal) leads to induced superconducting order, the same setup with (Bi$_{1-x}$Sb$_x$)$_2$Se$_3$ (another topological insulator) displays massive suppression of proximity pairing. On the other hand, in the class of systems studied here (low-dimensional OSMP iron-based materials), the pairing tendencies could arise from the intrinsic superconductivity of BaFe$_2$S$_3$ and BaFe$_2$Se$_3$ under pressure~\cite{Takahashi2015,Yamauchi2015,Ying2017} or doping~\cite{Patel2016,Patel2017}.

%--------------------------------------------------------------------------------
\begin{figure}[!htb]
\includegraphics[width=1.0\columnwidth]{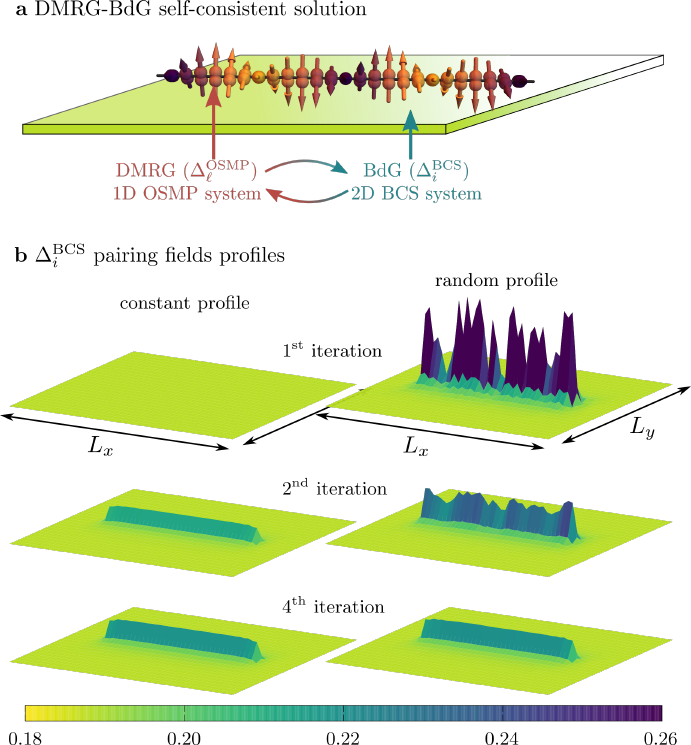}
\caption{{\bf BdG self-consistent solution.} {\bf a} Sketch of the hybrid DMRG-BdG algorithm. The OSMP chain is placed in the middle of a 2D BCS superconductor and coupled to it via the pairing field amplitudes present in both systems. {\bf b} Iteration dependence of the $\Delta^{\mathrm{BCS}}_{i}$ profiles for the case with the OSMP chain placed in the middle. Left (right) column depicts results initialized without (with random) pairing fields in the 1D OSMP system (calculated for $U/W=2$). See the Methods sections for details.}
\label{fig2}
\end{figure}
%--------------------------------------------------------------------------------

In order to keep our discussion general, we will make minimal assumptions on the SC state, and consider only the simplest on-site pairing. The reader should consider it either as the intrinsic pairing tendencies of the iron-based SC material or as the pairing field induced by the proximity to an $s$-wave SC substrate, e.g., Pb or Nb. Independently of its origin, the SC in the 1D OSMP system studied here must be investigated beyond the 1D lattice since the quantum fluctuations would inevitably destroy any long-range order. Therefore, let us first consider the OSMP chain placed atop the center of a two-dimensional (2D) BCS superconductor (see Fig.~\ref{fig2}a for a sketch) and the total system described by the Hamiltonian
\begin{equation}
H_\mathrm{tot}=H_\mathrm{gKH}+H_\mathrm{BCS} - V \sum_{\langle \ell, \ell^\prime \rangle} (c^{\dagger}_{\ell,\uparrow}c^{\dagger}_{\ell,\downarrow} a_{\ell^\prime,\downarrow} a_{\ell^\prime,\uparrow}+\mathrm{H.c.}).
\label{bdgcoup}
\end{equation}
Here, $\ell^\prime$ represents the single site within the 2D BCS system $H_\mathrm{BCS}$ which is closest to the site $\ell$ in the OSMP chain, and $a^{\phantom{\dagger}}_{i,\sigma},a^{\dagger}_{i,\sigma}$ stand for fermionic operators within the BCS superconductor (see the Methods section). The interaction between the subsystems [last term in Eq. (\ref{bdgcoup})] is studied within the BCS-like decoupling scheme, where we introduce the pairing amplitudes $\Delta^\mathrm{BCS}_{\ell^\prime}=\langle a_{\ell^\prime,\downarrow} a_{\ell^\prime,\uparrow}\rangle$ and $\Delta^\mathrm{OSMP}_\ell=\langle c_{\ell,\downarrow} c_{\ell,\uparrow}\rangle$ for the BCS superconductor and the OSMP chain, respectively. In order to fully take into account the many-body nature of the OSMP system, we have developed a hybrid algorithm, the details of which are given in the Methods section. In summary: we iteratively solve the OSMP chain and the BCS system by means of the DMRG and the Bogoliubov-de Gennes (BdG) equations, respectively. This back-and-forth computational setup is costly but important to gain confidence about our result.

We monitor the landscapes of pairing fields in both systems and exemplary results are presented in Fig.~\ref{fig2}b (for more results see the Supplementary Note~1). Initially, only the BCS system has finite, spatially uniform, pairing amplitudes $\Delta^\mathrm{BCS}_{\ell^\prime}$ (left column in Fig.~\ref{fig2}b), which are then used in the DMRG procedure applied to the OSMP Hamiltonian 
\begin{equation}
H=H_\mathrm{gKH}+\sum_{\ell}\Delta_\ell\left(c^{\dagger}_{\ell,\uparrow}c^{\dagger}_{\ell,\downarrow}+\mathrm{H.c.}\right),
\label{gKHSC}
\end{equation}
where $\Delta_\ell=- V\Delta^\mathrm{BCS}_{\ell^\prime}$. Next, the $\Delta^\mathrm{OSMP}_\ell$ set is calculated from DMRG and returned to the BdG equations relevant for the BCS system. The procedure is repeated until convergence is established. The results presented in Fig.~\ref{fig2}b show that already after $\sim4$ iterations the landscape of $\Delta^\mathrm{BCS}_{\ell^\prime}$ stabilizes to an interaction $U$ dependent value. We found that the resulting amplitude $\Delta_\ell=- V\Delta^\mathrm{BCS}_{\ell^\prime}$ is almost uniform within the OSMP chain. Furthermore, we have also confirmed that using extended $s$-wave pairing (creating pairs on nearest-neighbour sites) does not influence our conclusions. Therefore, in the remainder of the paper we use spatially uniform $\Delta_\ell=\Delta_\mathrm{SC}$ in Eq.~\eqref{gKHSC}. Also, in order to emphasize the role of interaction, in the main text we fix the pairing field to $\Delta_\mathrm{SC}/W\simeq0.5$. The detailed $\Delta_\mathrm{SC}$-dependence of our findings is discussed in the Supplementary Note~1.

%%%%%%%%%%%%%%%%%%%%%%%%%%%%%%%%%%%%%%%%%%%%%%%%%%%%%%%%%%%%%%%%%%%%%%%%%%%%%%%%%
\vspace{0.5em}
\noindent {\bf\uppercase{Results}}

\vspace{0.5em}
\noindent {\bf Magnetism of orbital-selective Mott phase} Previous work has shown that the OSMP (with $\Delta_\mathrm{SC}=0$) has a rich magnetic phase diagram~\cite{Herbrych2019-1}. (i) At small $U$ the system is paramagnetic. (ii) At $\overline{n}=1$ and $\overline{n}=0$ standard antiferromagnetic (AFM) order develops, $\uparrow\downarrow\uparrow\downarrow$, with total on-site magnetic moment $\mathbf{\langle S^2 \rangle}=S(S+1)=2$ and $3/4$, respectively. (iii) For $0<\overline{n}<1$ and $U\gg W$ the system is a ferromagnet (FM) $\uparrow\uparrow\uparrow\uparrow$. Interestingly, in the always challenging intermediate interaction regime $U\sim {\cal O}(W)$ the AFM- and FM-tendencies (arising from superexchange and double-exchange, respectively) compete and drive the system towards novel magnetic phases unique to multi-orbital systems. (iv) For $U\sim W$, the system develops a so-called block-magnetic order, consisting of FM blocks which are AFM coupled, e.g. $\uparrow\uparrow\downarrow\downarrow$, as sketched in Fig.~\ref{fig1}c. The block size appears controlled by the Fermi vector $k_{\mathrm{F}}$, i.e., the propagation wavevector of the block-magnetism is given by $q_{\mathrm{max}}=2k_{\mathrm{F}}$ (with $2k_{\mathrm{F}}=\pi\overline{n}$ for the chain lattice geometry). In this work we choose $\overline{n}=0.5$ (adjusted via the chemical potential $\mu$), as the relevant density for BaFe$_2$Se$_3$ $\pi/2$-block magnetic order~\cite{Mourigal2015}. Then, the latter order can be identified via the peak position of the static structure factor \mbox{$S(q)=\langle \mathbf{T}_{-q}\cdot\mathbf{T}_{q}\rangle$} at $q_{\mathrm{max}}=\pi/2$ or via a finite dimer order parameter \mbox{$D_{\pi/2}=\sum_\ell (-1)^\ell \langle \mathbf{T}_{\ell}\cdot \mathbf{T}_{\ell+1}\rangle/L$}, where we introduced the Fourier transform $\mathbf{T}_{q}=\sum_\ell \exp(i q \ell)\,\mathbf{T}_\ell/\sqrt{L}$ of the total spin operator $\mathbf{T}_{\ell}=\mathbf{S}_{\ell}+\mathbf{s}_{\ell}$. In Fig.~\ref{fig1}c $S(q)$ is shown at moderate interaction: at $U/W<1.6$ it displays a maximum at $q_{\mathrm{max}}=\pi/2$, consistent with $\uparrow\uparrow\downarrow\downarrow$ order.

%--------------------------------------------------------------------------------
\begin{figure*}[!htb]
\includegraphics[width=1.0\textwidth]{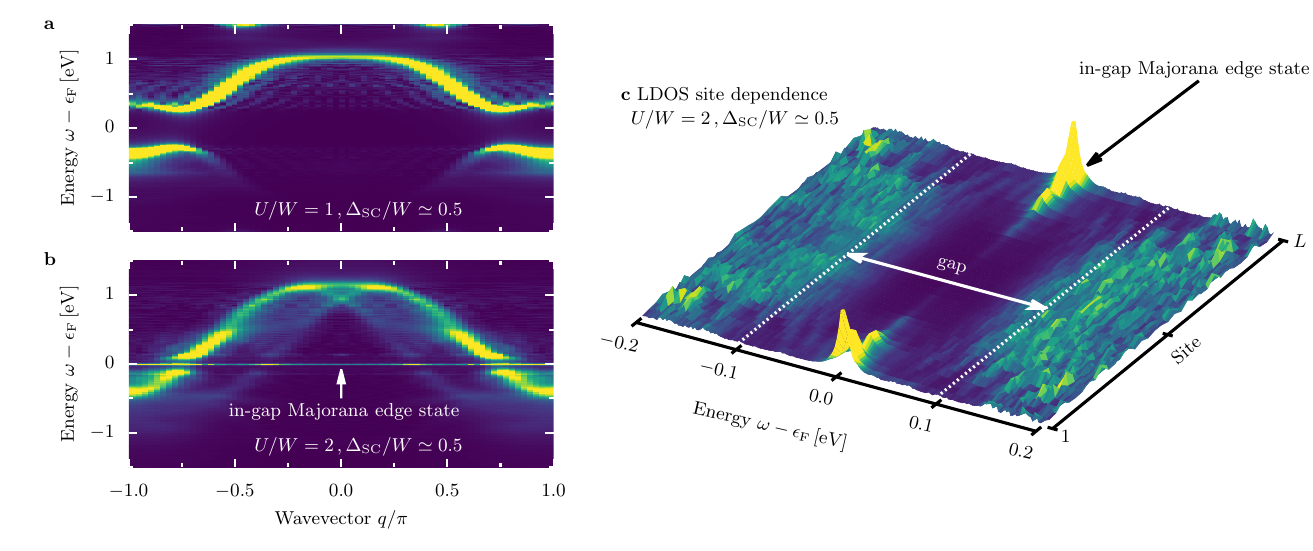}
\caption{{\bf Spectral functions}. Effect of the finite pairing field $\Delta_{\mathrm{SC}}/W\simeq 0.5$ on the single-particle spectral function $A(q,\omega)$ for {\bf a} $U/W=1$ and {\bf b} $U/W=2$ calculated for $L=36$, $\overline{n}=0.5$, and $\delta\omega=0.02\,\mathrm{[eV]}$. Majorana zero-energy modes are indicated in {\bf b}. {\bf c} Local density-of-states (LDOS) in the in-gap frequency region ($\delta\omega=0.002\,\mathrm{[eV]}$) vs chain site index. The sharp LDOS peaks at the edges represent Majorana edge states, while the bulk of the system exhibits gapped behaviour.}
\label{fig3}
\end{figure*}
%--------------------------------------------------------------------------------

Remarkably, it has been shown recently~\cite{Herbrych2019-2} that there exists an additional unexpected phase in between the block- and FM-ordering. Namely, upon increasing the interaction ($1.6<U/W<2.4$), the maximum of $S(q)$ in Fig.~\ref{fig1}c shifts towards incommensurate wavevectors (while for $U/W>2.4$ the system is a ferromagnet). This incommensurate region reflects a novel magnetic spiral where the magnetic islands maintain their ferromagnetic character (with $D_{\pi/2}\ne0$) but start to rigidly rotate, forming a so-called block-spiral (see sketch Fig.~\ref{fig1}c). The latter can be identified by a large value~\cite{Herbrych2019-2} of the long-range chirality correlation function $\langle \boldsymbol{\kappa}_{\ell} \cdot \boldsymbol{\kappa}_{m}\rangle$ where $\boldsymbol{\kappa}_{\ell}=\mathbf{T}_{\ell}\times\mathbf{T}_{\ell+N}$ and $N$ is the block size. It is important to note that the spiral magnetic order appears without any direct frustration in the Hamiltonian \eqref{hamkon}, but rather is a consequence of hidden frustration caused by competing energy scales in the OSMP regime. Finally, it should be noted that the block-spiral OSMP state is not limited to one-dimensional chains. In the Supplementary Note~2, we show similar investigations for the ladder geometry and find rigidly rotating $2\times 2$ FM islands. These results are consistent with recent nuclear magnetic resonance measurements on the CsFe$_2$Se$_3$ ladder compound which reported the system's incommensurate ordering~\cite{Murase2020}.

Interestingly, an interaction induced spiral order is also present when  SC pairing is included in the model, as evident form Fig.~\ref{fig1}d. However, the spiral mutates from block- to canonical-type with $D_{\pi/2}=0$ (see sketch in Fig.~\ref{fig1}d), indicating an unusual back-and-forth feedback between magnetism and superconductivity. As discussed below, the pairing optimizes the spiral profile to properly create the Majoranas. The competition between many energy scales (Hubbard interaction, Hund exchange, and SC pairing) leads to a novel phenomena: an interaction induced topological phase transition into a many-body state with MZM, unconventional SC, and canonical spiral.

%%%%%%%%%%%%%%%%%%%%%%%%%%%%%%%%%%%%%%%%%%%%%%%%%%%%%%%%%%%%%%%%%%%%%%%%%%%%%%%%%
\vspace{0.5em}
\noindent {\bf Majorana fermions} Figure~\ref{fig3} shows the effect of $\Delta_{\mathrm{SC}}/W\simeq 0.5$ on the single-particle spectral function $A(q,\omega)$ (see Methods section) for the two crucial phases in our study, the block-collinear and block-spiral magnetic orders ($U/W=1$ and $U/W=2$, respectively). As expected, in both cases, a finite superconducting gap opens at the Fermi level $\epsilon_{\mathrm{F}}$ ($\sim 0.5$~[eV] for $U/W=1$ and $\sim 0.1$~[eV] for $U/W=2$). Remarkably, in the block-spiral phase an additional prominent feature appears: a sharply localized mode inside the gap at $\epsilon_{\mathrm{F}}$, displayed in Fig.~\ref{fig3}b. Such an in-gap mode is a characteristic feature of a topological state, namely the bulk of the system is gapped, while the edge of the system contains the in-gap modes. To confirm this picture, in Fig.~\ref{fig3}c, we present a high-resolution frequency data of the real-space local density-of-states (LDOS; see Methods Section) near the Fermi energy $\epsilon_{\mathrm{F}}$. As expected, for the topologically nontrivial phase, the zero energy modes are indeed confined to the system's edges. It is important to note that this phenomenon is absent for weaker interaction $U/W=1$. Furthermore, one cannot deduce this behaviour from the $U\to\infty$ or $J_{\mathrm{H}}\to \infty$ limits, where the system has predominantly collinear AFM or FM ordering, leading again to a trivial superconducting behaviour. However, as shown below, at moderate $U$ the competing energy scales present in the OSMP lead to the topological phase transition controlled by the electron-electron interaction.

\begin{figure}[!htb]
\includegraphics[width=1.0\columnwidth]{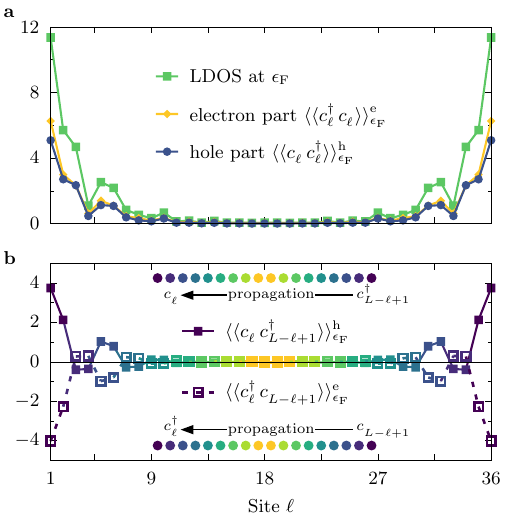}
\caption{{\bf Correlation functions of Majorana fermions.} {\bf a} Site $\ell$ dependence of the local density-of-states (LDOS) at the Fermi level ($\omega=\epsilon_{\mathrm{F}}$) together with its hole $\langle\langle c^{\phantom{\dagger}}_{\ell}\, c^{\dagger}_{\ell} \rangle\rangle_{\epsilon_{\mathrm{F}}}^{\mathrm{h}}$ and electron $\langle\langle c^{\dagger}_{\ell} \, c^{\phantom{\dagger}}_{\ell} \rangle\rangle_{\epsilon_{\mathrm{F}}}^{\mathrm{e}}$ contributions. {\bf b} Site dependence of the centrosymmetric spectral function $\langle\langle c^{\phantom{\dagger}}_{\ell}\,c^{\dagger}_{L-\ell+1} \rangle\rangle_{\epsilon_{\mathrm{F}}}^{\mathrm{h}}$ and $\langle\langle c^{\dagger}_{\ell}\,c^{\phantom{\dagger}}_{L-\ell+1} \rangle\rangle_{\epsilon_{\mathrm{F}}}^{\mathrm{e}}$. Sketches represent the calculated process: the probability of creating the electron on one end of the system (site $\ell$) and a hole at the opposite end (site $L-\ell+1$), or vice-versa, at given energy $\omega$. The pairs of sites where the spectral function is evaluated are represented by the same colors. All results were calculated for $L=36$, $U/W=2$, $\Delta_{\mathrm{SC}}/W\simeq 0.5$, and $\overline{n}=0.5$.}
\label{fig4}
\end{figure}

Let us now identify the induced topological state. The site-dependence of the local density-of-states presented in Fig.~\ref{fig3}c reveals zero-energy edge modes, namely peaks at frequency $\omega \simeq \epsilon_{\mathrm{F}}$ localized at the edges of the chain with open ends. While such modes are a characteristic property of the MZM, finding peaks in the LDOS alone is insufficient information for an unambiguous identification. To demonstrate that the gKH model with superconductivity indeed hosts Majorana modes, we have numerically checked three distinct features of the MZM: 

\noindent (i) Since the Majorana particles are their own antiparticles, the spectral weight of the localized modes should be built on an equal footing from the electron and hole components. Figure~\ref{fig4}a shows that this is indeed the case. 

\begin{figure}[!htb]
\includegraphics[width=1.0\columnwidth]{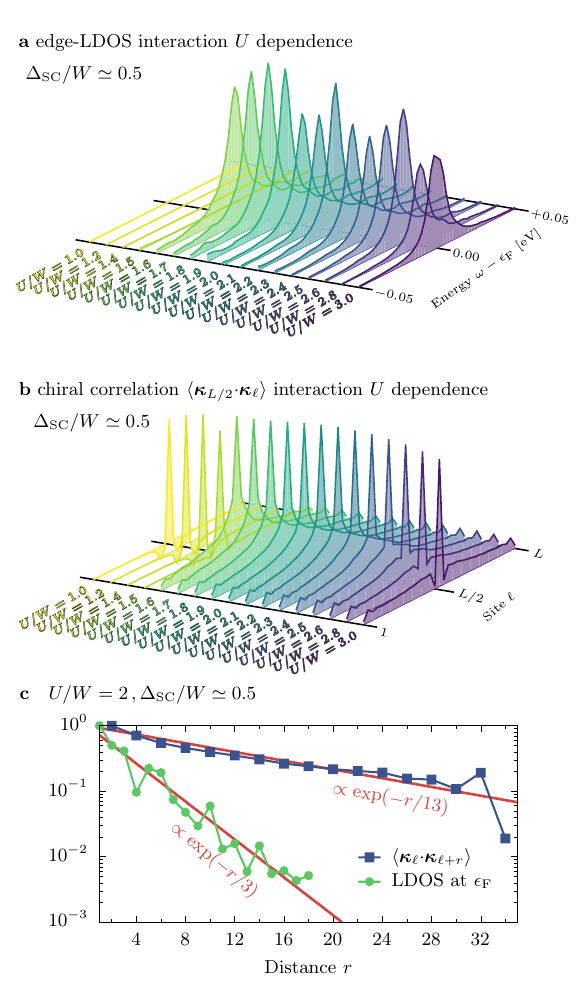}
\caption{{\bf Interaction dependence of the MZM.} Dependence on the Hubbard interaction $U$ of {\bf a} the edge-LDOS at site $\ell=1$ (near the Fermi level $\epsilon_{\mathrm{F}}$) and {\bf b} the chirality correlation function $\langle \boldsymbol{\kappa}_{L/2} \cdot \boldsymbol{\kappa}_{\ell}\rangle$. All results calculated for $\Delta_{\mathrm{SC}}/W\simeq 0.5$, $\overline{n}=0.5$, $L=36$. {\bf c} Spatial decay of the local density-of-states at the Fermi level (LDOS at $\epsilon_{\mathrm{F}}$) and the chirality correlation function $\langle \boldsymbol{\kappa}_{\ell} \cdot \boldsymbol{\kappa}_{\ell+r}\rangle$ for $U/W=2$. Red solid lines indicate exponential decay $\exp(-r/l_{\alpha})$ with $l_{\mathrm{MZM}}=3$ and $l_{\mathrm{s}}=15$, for the MZM and the spiral order, respectively.}
\label{fig5}
\end{figure}

\noindent (ii) The total spectral weight present in the localized modes can be rigorously derived from the assumption of the MZM's existence (see Methods section), and it should be equal to $0.5$. Integrating our DMRG results in Fig.~\ref{fig3}c over a narrow energy window and adding over the first few edge sites gives $\simeq0.47$, very close to the analytical prediction. Note that the Majoranas are not strictly localized at one edge site $\ell\in\{1,L\}$, as evident from Fig.~\ref{fig4}a. Instead, the MZM are exponentially decaying over a few sites (see Fig.~\ref{fig5}c), and we must add the spectral weight accordingly (separately for the left and right edges). 

%--------------------------------------------------------------------------------
%--------------------------------------------------------------------------------

\noindent (iii) The MZM located at the opposite edges of the system form one fermionic state, namely the edge MZM are correlated with one another over large distances. To show such a behaviour, consider the hole- and electron-like centrosymmetric spectral functions, $\langle\langle c^{\phantom{\dagger}}_{\ell}\, c^{\dagger}_{L-\ell+1} \rangle\rangle_{\omega}^{\mathrm{h}}$ and $\langle\langle c^{\dagger}_{\ell} \, c^{\phantom{\dagger}}_{L-\ell+1} \rangle\rangle_{\omega}^{\mathrm{e}}$, respectively. These functions represent the probability amplitude of creating an electron on one end and a hole at the opposite end (or vice-versa) at a given energy $\omega$ (see Methods section for detailed definitions and Supplementary Note~3 for further discussion). Figure~\ref{fig4}b shows $\langle\langle c^{\phantom{\dagger}}_{\ell}\, c^{\dagger}_{L-\ell+1} \rangle\rangle_{\omega}^{\mathrm{h}}$ and $\langle\langle c^{\dagger}_{\ell} \, c^{\phantom{\dagger}}_{L-\ell+1} \rangle\rangle_{\omega}^{\mathrm{e}}$ at the Fermi level $\omega=\epsilon_{\mathrm{F}}$, namely in the region where the MZM should be present. As expected, the bulk of the system behaves fundamentally different from the edges. In the former, crudely when $L/2\lesssim\ell\lesssim 3L/4$, the aforementioned spectral functions vanish reflecting the gapped (bulk) spectrum with lack of states at the Fermi level. However, at the boundaries ($\ell\ll L/2$ and $\ell\gg L/2$) the values of the centrosymmetric spectral functions are large, with maximum at the edges $\ell\in\{1,L\}$. The long-range (across the system) correlations of the edge states strongly support their topological nature.

%--------------------------------------------------------------------------------
%--------------------------------------------------------------------------------

Finally, let us discuss the physical mechanism causing the onset of MZM. In Fig.~\ref{fig5}a we present the Hubbard $U$ interaction dependence of the edge-LDOS ($\ell=1$) in the vicinity of the Fermi level, $\omega\sim\epsilon_{\mathrm{F}}$. It is evident from the presented results that the edge-LDOS acquires a finite value quite abruptly for $U>U_{\mathrm{c}}\simeq1.5$. To further clarify this matter, let us return to the magnetic states in the OSMP regime. Figure~\ref{fig5}b shows the chirality correlation function $\langle \boldsymbol{\kappa}_{L/2} \cdot \boldsymbol{\kappa}_{\ell}\rangle$ (with \mbox{$\boldsymbol{\kappa}_\ell=\mathbf{T}_\ell\times\mathbf{T}_{\ell+1}$}) for increasing value of the Hubbard $U$ strength. We observe a sudden appearance of the chirality correlation exactly at $U_{\mathrm{c}}$, a behaviour similar to that of the edge LDOS. Interestingly, in the system without the pairing field, $\Delta_{\mathrm{SC}}=0$, at a similar value of $U\simeq1.6$ the system enters the block-spiral phase with rigidly rotating FM islands. However, in our setup, the tendencies of OSMP to create magnetic blocks \cite{Herbrych2019-1} are highly suppressed by empty and doubly occupied sites favored by the finite pairing field $\Delta_{\mathrm{SC}}$. As a consequence, the block-spiral order is reshaped to a canonical type of spiral without dimers $D_{\pi/2}=0$. This behaviour is similar to the MZM observed when combining $s$-wave SC with a classical magnetic moments heterostructure~\cite{Braunecker2013,Perge2013,Klinovaja2013,Vazifeh2013}. In the latter, the Ruderman–Kittel–Kasuya–Yosida (RKKY) mechanism stabilizes a classical long-range spiral with $2k_{\mathrm{F}}$ pitch (where $k_{\mathrm{F}}\propto\overline{n}$ is the Fermi wavevector). Within the OSMP, however, the pitch is, on the other hand, controlled by the interaction $U$ (at fixed $\overline{n}$), as evident from the results presented in Figs.~\ref{fig1}b and \ref{fig1}c. 

Furthermore, analysis of the chirality correlation function $\langle \boldsymbol{\kappa}_{\ell} \cdot \boldsymbol{\kappa}_{\ell+r}\rangle$ indicates that the spiral order decays with the distance $r$ (see Fig.~\ref{fig5}c), as expected in a 1D quantum system. Note, however, that the MZM decay length scale, $l_{\mathrm{MZM}}$, and that of the spiral, $l_\mathrm{s}$, differ substantially. The Majoranas are predominantly localized at the system edges, thus yielding a short localization length $l_{\mathrm{MZM}}\simeq3$. The spiral, although still decaying exponentially, has a robust correlation length $l_\mathrm{s}\simeq13$, of the same magnitude as the $\Delta_{\mathrm{SC}}=0$ result~\cite{Herbrych2019-2}. Then, for large but finite chains the overlap between the edge modes is negligible while the magnetic correlations on the distance $L$ are still large enough to promote triplet pairing and the Majorana modes. In addition, we have observed that smaller values of $\Delta_\mathrm{{SC}}$ than considered here also produce the MZM. However, since the Majoranas have an edge localization length inversely proportional to $\Delta_{\mathrm{SC}}$, reducing the latter leads to an overlaps between the left and right Majorana states in our finite systems \cite{Kitaev2001,Stanescu2013}, thus distorting the physics we study. After exploration, $\Delta_\mathrm{SC}/W\simeq0.5$ was considered an appropriate compromise to address qualitatively the effects of our focus given our practical technical constraints within DMRG (see Supplementary Note~4 for details).

Conceptually, it is important to note that the interaction-induced spiral at $U/W=2$ is not merely frozen when $\Delta_{\mathrm{SC}}$ increases. Specifically, the characteristics \cite{Herbrych2019-2} of the chirality correlation function $\langle \boldsymbol{\kappa}_{i} \cdot \boldsymbol{\kappa}_{j}\rangle$ qualitatively differ beween the trivial ($\Delta_{\mathrm{SC}}=0$) and topological phases ($\Delta_{\mathrm{SC}}\ne0$): increasing $\Delta_{\mathrm{SC}}$ suppresses the dimer order and leads to a transformation from block spiral to a standard canonical spiral with $D_{\pi/2}=0$ in the topologically nontrivial phase. As a consequence, the proximity to a superconductor influences on the magnetic order to optimize the spin pattern needed for MZM. Surprisingly, $\Delta_\mathrm{SC}$ influences on the collinear spin order as well. In fact, at $U/W=1$, before spirals are induced, the proximity to superconductivity changes the block spin order into a more canonical staggered spin order to optimize the energy (see Fig.~\ref{fig3}b). This is a remarkable, and unexpected, back-and-forth positive feedback between degrees of freedom that eventually causes the stabilization of the MZM.

%--------------------------------------------------------------------------------
\begin{figure}[!htb]
\includegraphics[width=1.0\columnwidth]{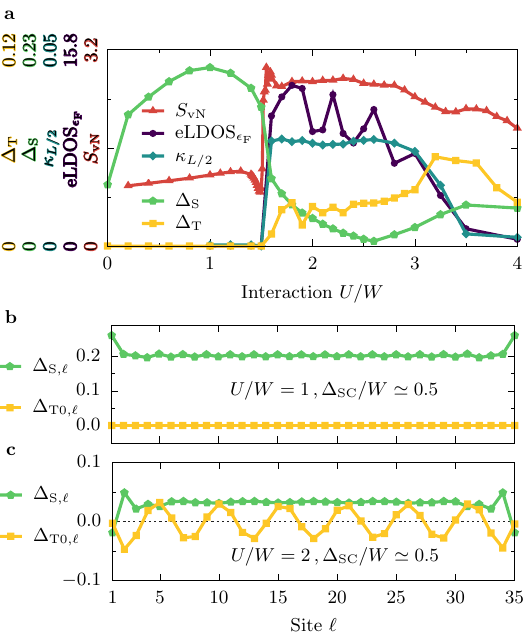}
\caption{{\bf Phase diagram.} {\bf a} Hubbard interaction $U$ dependence of the (i) von Neumann entanglement entropy $S_\mathrm{vN}$, (ii) edge local density-of-states at the Fermi level (eLDOS$_{\epsilon_\mathrm{F}}$), (iii) the value of chirality correlation function at distance $L/2$ (i.e. $\langle \boldsymbol{\kappa}_{L/4} \cdot \boldsymbol{\kappa}_{3L/4}\rangle$), as well as (iv) nonlocal singlet $\Delta_{\mathrm{S}}$ and triplet $\Delta_{\mathrm{T}}$ pairing amplitudes. See text for details. All results were calculated for $L=36$, $\Delta_{\mathrm{SC}}/W\simeq 0.5$, and $\overline{n}=0.5$. {\bf b-c} Spatial dependence of the singlet and triplet SC amplitudes, $\Delta_{\mathrm{S},\ell}$ and $\Delta_{\mathrm{T0},\ell}$, respectively (see Methods section for details), with {\bf b} the trivial phase ($U/W=1\,,\Delta_{\mathrm{SC}}/W\simeq 0.5$) and {\bf c} the topological phase ($U/W=2\,,\Delta_{\mathrm{SC}}/W\simeq 0.5$). In the latter, the oscillations of the triplet component are related to the pitch of the underlying spiral magnetic order.}
\label{fig6}
\end{figure}
%--------------------------------------------------------------------------------

%%%%%%%%%%%%%%%%%%%%%%%%%%%%%%%%%%%%%%%%%%%%%%%%%%%%%%%%%%%%%%%%%%%%%%%%%%%%%%%%%
\vspace{0.5em}
\noindent{\bf\uppercase{Discussion}}

\noindent Our main findings are summarized in Fig.~\ref{fig6}: upon increasing the strength of the Hubbard interaction $U$ within the OSMP with added SC pairing field, the system undergoes a topological phase transition. The latter can be detected as the appearance of edge modes which are mutually correlated in a finite system. This in turn leads to, e.g., the sudden increase of the entanglement, as measured by the von Neumann entanglement entropy $S_{\mathrm{vN}}$ (see the Methods section). The transition is driven by the change in the magnetic properties of the system, namely by inducing a finite chirality visible in the correlation function $\langle \boldsymbol{\kappa}_{\ell} \cdot \boldsymbol{\kappa}_{m}\rangle$. The above results are consistent with the appearance of the MZM at the topological transition. It should be noted that the presence of those MZM implies unconventional $p$-wave superconductivity~\cite{Pawlak2019}. As a consequence, for our description to be consistent, the topological phase transition ought to be accompanied by the onset of triplet SC amplitudes $\Delta_{\mathrm{T}}$. To test this nontrivial effect, we monitored the latter, together with the singlet SC amplitude $\Delta_{\mathrm{S}}$ (related to a nonlocal $s$-wave SC; see the Methods section for detailed definitions). As is evident from the results in Fig.~\ref{fig6}, for $U<U_{\mathrm{c}}$ we observe only the singlet component $\Delta_{\mathrm{S}}$ canonical for an $s$-wave SC, while for $U\geq U_{\mathrm{c}}$ the triplet amplitude $\Delta_{\mathrm{T}}$ develops a robust finite value. It is important to stress that $\Delta_{\mathrm{T}}\ne0$ is an emergent phenomenon, induced by the correlations present within the OSMP, and is not assumed at the level of the model (we use a trivial on-site $s$-wave pairing field as input).

In summary, we have shown that the many competing energy scales induced by the correlation effects present in superconducting multi-orbital systems within OSMP lead to a topological phase transition. Differently from the other proposed MZM candidate setups, our scheme does not require frozen classical magnetic moments or, equivalently, FM ordering in the presence of the Rashba spin-orbit coupling \cite{Braunecker2010}. All ingredients necessary to host Majorana fermions appear as a consequence of the quantum effects induced by the electron-electron interaction. The pairing filed can be induced by the proximity effect with a BCS superconductor, or it could be an intrinsic property of some iron superconductors under pressure or doping. It is important to note that the coexistence of SC and nontrivial magnetic properties is mostly impossible in single-orbital systems. Here, the OSMP provides a unique platform in which this constraint is lifted by, on the one hand, spatially separating such phenomena, and, on the other hand, strongly correlating them with each other. Furthermore, our proposal allows to study the effect of quantum fluctuations on the MZM modes. There are only a few candidate materials that may exhibit the behaviour found here. The block-magnetism (a precursor of the block-spiral phase) was recently argued to be relevant for the chain compound Na$_2$FeSe$_2$~\cite{Pandey2020}, and was already experimentally found in the BaFe$_2$Se$_3$ ladder~\cite{Mourigal2015}. Incommensurate order was reported in CsFe$_2$Se$_3$~\cite{Murase2020}. Also, the OSMP~\cite{Caron2012,Ootsuki2015,Craco2020} and superconductivity~\cite{Takahashi2015,Yamauchi2015,Ying2017} proved to be important for other compounds from the 123 family of iron-based ladders.

Our findings provide also a new perspective to the recent reports of topological superconductivity and Majorana fermions found in two-dimensional compounds Fe(Se,Te)~\cite{Wang2018,Zhang2018,Machida2019,Wang2020,Chen2020}. Since orbital-selective features were observed in clean FeSe~\cite{Yu2018,Kostin2018}, it is reasonable to assume that OSMP is also relevant for doped Fe(Se,Te)~\cite{Jiang2020}. Regarding magnetism, the ordering of FeSe was mainly studied within the classical long-range Heisenberg model~\cite{Glasbrenner2015}, where block-like structures (e.g., double stripe or staggered dimers) dominate the phase diagram for realistic values of the system parameters. Note that the effective spin model of the block-spiral phase studied here was also argued to be long-ranged~\cite{Herbrych2019-2}. The aforementioned phases of FeSe are typically neighboring (or are even degenerate with) the frustrated spiral-like magnetic orders~\cite{Glasbrenner2015}, also consistent with the OSMP magnetic phase diagram~\cite{Herbrych2019-1}. In view of our results, the following rationale could be used to explain the behaviour of the above materials: the competing energy scales present in multi-orbital iron-based compounds, induced by changes in the Hubbard interaction due to chemical substitution or pressure, lead to exotic magnetic spin textures. The latter, together with the superconducting tendencies, lead to topologically nontrivial phases exhibiting the MZM~\cite{Nakosai2013,Steffensen2020}. Also, similar reasonings can be applied to the heavy-fermion metal UTe$_2$. It was recently shown that this material displays spin-triplet superconductivity~\cite{Jiao2020} together with incommensurate magnetism~\cite{Duan2020}.

%%%%%%%%%%%%%%%%%%%%%%%%%%%%%%%%%%%%%%%%%%%%%%%%%%%%%%%%%%%%%%%%%%%%%%%%%%%%%%%%%
\vspace{0.5em}
\noindent{\bf\uppercase{Methods}}

%%%%%%%%%%%%%%%%%%%%%%%%%%%%%%%%%%%%%%%%%%%%%%%%%%%%%%%%%%%%%%%%%%%%%%%%%%%%%%%%%
\vspace{0.5em}
\noindent {\bf DMRG method.} 
The Hamiltonians and observables discussed here were studied using the density matrix renormalization group (DMRG) method~\cite{white1992,schollwock2005} within the single-center site approach~\cite{white2005}, where the dynamical correlation functions are evaluated via the dynamical-DMRG~\cite{jeckelmann2002,nocera2016}, i.e., calculating spectral functions directly in frequency space with the correction-vector method using Krylov decomposition~\cite{nocera2016}. We have kept up to $M=1200$ states during the DMRG procedures, allowing us to accurately simulate system sizes up to $L=48$ and $L=60$ with truncation errors $\sim10^{-8}$ and $\sim10^{-6}$, respectively. 

We have used the \textsc{DMRG++} computer program developed at Oak Ridge National Laboratory (\href{https://g1257.github.io/dmrgPlusPlus/}{https://g1257.github.io/dmrgPlusPlus/}). The input scripts for the \textsc{DMRG++} package to reproduce our results can be found at \href{https://bitbucket.org/herbrychjacek/corrwro/}{https://bitbucket.org/herbrychjacek/corrwro/} and also on the \textsc{DMRG++} package webpage.

%%%%%%%%%%%%%%%%%%%%%%%%%%%%%%%%%%%%%%%%%%%%%%%%%%%%%%%%%%%%%%%%%%%%%%%%%%%%%%%%%
\vspace{0.5em}
\noindent {\bf Hybrid DMRG-BdG algorithm.}
We consider a 2D, $s$-wave, BCS superconductor at half-filling,
\begin{eqnarray}
H_\mathrm{BCS}=&-&t_\mathrm{BCS}\sum_{\langle i,j\rangle,\sigma} a^{\dagger}_{i,\sigma} a^{\phantom{\dagger}}_{j,\sigma}\nonumber\\
&-&V_\mathrm{BCS}\sum_i \left(\Delta^\mathrm{BCS}_i\,a^{\dagger}_{i,\uparrow}a^{\dagger}_{i,\downarrow}+\mathrm{H.c.}\right).
\label{hbcs}
\end{eqnarray}
Here $\langle i,j\rangle$ denotes summation over nearest-neighbour sites of a square lattice and $a^{\dagger}_{i,\sigma}$ ($a^{\phantom{\dagger}}_{i,\sigma}$) creates (destroys) an electron with spin projection $\sigma=\{\uparrow,\downarrow\}$ at site $i$. The BCS system is coupled to the OSMP chain, as described by the last term of Hamiltonian (\ref{bdgcoup}) in the main text. At the BCS level, the latter term emerges as additional (external) pairing field to the OSMP region
\begin{eqnarray}
H_V=&-&V \sum_{\ell^\prime} \left(\Delta^\mathrm{OSMP}_{\ell}\,a^{\dagger}_{\ell^\prime,\uparrow}a^{\dagger}_{\ell^\prime,\downarrow}+\mathrm{H.c.}\right).
\label{hv}
\end{eqnarray}
Here, the summation is restricted to the sites of the BCS system which are coupled to the OSMP chain. In numerical calculations, we set the hopping integral $t_\mathrm{BCS}=2\,\mathrm{[eV]}$, fix the system size to $L_x=54$ and $L_y=27$ (with 1D OSMP system coupled to the $\ell^\prime=14$ row of sites), use the BCS attractive potential $V_\mathrm{BCS}/t_\mathrm{BCS}=2$ and the coupling strength $V/t_\mathrm{BCS}=2$. Although we assume periodic boundary conditions for the BCS system, the translational invariance is broken by the coupling to the OSMP chain.

Our procedure consists of two alternating steps:\\
\noindent 1. BdG calculations. In the first step we assume an initial set $\Delta^\mathrm{OSMP}_{\ell}$ and diagonalize the Hamiltonian $H_\mathrm{BCS}+H_V$, as defined in Eqs. (\ref{hbcs}) and (\ref{hv}). To this end, we use the standard BdG equations at zero temperature. They yield self-consistent results for the pairing amplitude, $\Delta^\mathrm{BCS}_i=\langle a_{i,\downarrow} a_{i,\uparrow} \rangle$, for all sites $i$ within the BCS system. From among the latter results, we single out the amplitudes $\Delta^\mathrm{BCS}_{\ell^\prime}$ on the sites $i={\ell^\prime}$ which are coupled to the OSMP chain. \\
\noindent 2. DMRG calculations. The OSMP system with \mbox{$\Delta_\ell=-V \Delta^\mathrm{BCS}_{\ell^\prime}$} in Eq.~\eqref{gKHSC} is evaluated using the DMRG approach. The spatially dependent amplitudes $\Delta^\mathrm{OSMP}_\ell$ are calculated providing a new set of external fields for the subsequent BdG calculations.

The above procedure is repeated iteratively until we obtain converged results. In the main text (see Fig.~\ref{fig2}) we presented results of the above algorithm starting from $\Delta^\mathrm{OSMP}_{\ell}=0$. However, the procedure can also start from arbitrary pairing fields $\Delta^\mathrm{OSMP}_\ell$ in the first step. The right column of Fig.~\ref{fig2}b depicts results obtained using a random initial profile $\Delta^\mathrm{OSMP}_\ell\in[0,1]$. It is evident from the presented results that the converged result is independent from the initial configuration of $\Delta^\mathrm{OSMP}_\ell$ (at least for the couplings studied here). See the Supplementary Note~1 for further discussion and additional results.

%%%%%%%%%%%%%%%%%%%%%%%%%%%%%%%%%%%%%%%%%%%%%%%%%%%%%%%%%%%%%%%%%%%%%%%%%%%%%%%%%
\vspace{0.5em}
\noindent {\bf Spectral functions.}
Let us define the site-resolved frequency $(\omega)$ dependent electron (e) and hole (h) correlation functions
\begin{equation}
\langle\langle A_{\ell}\,B_{m} \rangle\rangle_{\omega}^{\mathrm{e},\mathrm{h}}=-\frac{1}{\pi}\mathrm{Im}\langle \mathrm{gs}|A_{\ell}\frac{1}{\omega^{+}\mp(H-\epsilon_0)}B_{m}|\mathrm{gs}\rangle\,,
\end{equation}
where the signs $+$ and $-$ should be taken for $\langle\langle ... \rangle\rangle_{\omega}^{e}$ and $\langle\langle \dots \rangle\rangle_{\omega}^{h}$, respectively. Here, $|\mathrm{gs}\rangle$ is the ground-state, $\epsilon_0$ the ground-state energy, and $\omega^{+}=\omega+i\eta$ with $\eta$ a Lorentzian-like broadening. For all results presented here we choose $\eta=2\delta\omega$, with $\delta\omega/W=0.001$ (unless stated otherwise).

The single-particle spectral functions $A(q,\omega)=A^{\mathrm{e}}(q,\omega)+A^{\mathrm{h}}(q,\omega)$, where $A^{\mathrm{e}}$ ($A^{\mathrm{h}}$) represent the electron (hole) part of the spectrum, have a standard definition, 
\begin{eqnarray}
A^{\mathrm{h}}(q,\omega)&=&\sum_{\ell} \mathrm{e}^{-iq(\ell-L/2)} \,\langle\langle c^{\phantom{\dagger}}_{\ell}\,c^{\dagger}_{L/2} \rangle\rangle_{\omega}^{\mathrm{h}}\,,\nonumber\\
A^{\mathrm{e}}(q,\omega)&=&\sum_{\ell} \mathrm{e}^{+iq(\ell-L/2)} \,\langle\langle c^{\dagger}_{\ell}\,c^{\phantom{\dagger}}_{L/2} \rangle\rangle_{\omega}^{\mathrm{e}}\,,
\label{akwdef}
\end{eqnarray}
with $c^{\phantom{\dagger}}_{\ell}=\sum_{\sigma}c^{\phantom{\dagger}}_{\ell,\sigma}$. Finally, the local density-of-states is defined as
\begin{equation}
\mathrm{LDOS}(\ell,\omega)
=\langle\langle c^{\phantom{\dagger}}_{\ell}\,c^{\dagger}_{\ell} \rangle\rangle_{\omega}^{\mathrm{h}}
+\langle\langle c^{\dagger}_{\ell}\,c^{\phantom{\dagger}}_{\ell} \rangle\rangle_{\omega}^{\mathrm{e}}\,.
\label{ldos}
\end{equation}

%%%%%%%%%%%%%%%%%%%%%%%%%%%%%%%%%%%%%%%%%%%%%%%%%%%%%%%%%%%%%%%%%%%%%%%%%%%%%%%%%
\vspace{0.5em}
\noindent {\bf Spectral functions of Majorana edge-states.} 
For simplicity, in this section we suppress the spin index $\sigma$ and assume that the lattice index $j$ contains all local quantum numbers. The many-body Hamiltonian is originally expressed in terms of fermionic operators $c^{\left( \dagger \right)}_j$, but it may be equivalently rewritten using the Majorana fermions (not to be confused with the MZM):
\begin{equation}
\gamma_{2j-1}=c^{\phantom{\dagger}}_{j}+c^{\dagger}_j,\quad \quad \gamma_{2j}=-i (c^{\phantom{\dagger}}_{j}-c^{\dagger}_j)\,,
\label{trans1}
\end{equation}
where $\gamma^{\dagger}_{l}=\gamma_{l}$ and $\{\gamma_{i},\gamma_{j}\}=2\delta_{ij}$. The latter anticommutation relation is invariant under orthogonal transformations, thus we can rotate the Majorana fermions arbitrarily with
\begin{equation}
\Gamma_{a}=\sum_{j} {\hat V}_{aj} \gamma_{j}\,,
\label{trans2} 
\end{equation}
where $\hat V$ are real, orthogonal matrices $\hat V^{\top} \hat V= \hat V \hat V^{\top} =1$. If the system hosts a pair of Majorana edge modes, $\Gamma_L$ and $\Gamma_R$, then we can find a transformation $\hat V$ such that the following Hamiltonian captures the low-energy physics
\begin{equation} 
H \simeq \,i \frac{\varepsilon}{2}\,\Gamma_L \, \Gamma_R + H'\,.
\label{heff}
\end{equation} 
It is important to note that $H'$ does not contribute to the in-gap states. It contains all Majorana operators, $\Gamma_a$, other than the MZM ($\Gamma_L$ and $\Gamma_R$). The first term in Eq.~\eqref{heff} arises from the overlap of the MZM in a finite system, while in the thermodynamic limit $\varepsilon \to 0$ both $\Gamma_L$ and $\Gamma_R$ become strictly the zero modes. While the ground state properties obtained from the zero temperature DMRG do not allow us to formally construct the transformation $\hat V$, we demonstrate below that the computed local and non-local spectral functions are fully consistent with the MZM. In fact, we are not aware of any other scenario that could explain the spectral functions reported in this work.

Let us investigate the retarded Green's functions
\begin{eqnarray}
G^{\mathrm{h}}\left(c^{\phantom{\dagger}}_j,c^{\dagger}_l \right) & = & -i\int^{\infty}_0 {\rm d}t\, e^{i \omega t} \langle \mathrm{gs} | c^{\phantom{\dagger}}_j(t) c^{\dagger}_l | \mathrm{gs} \rangle\,, \nonumber \\ 
G^{\mathrm{e}}\left(c^{\phantom{\dagger}}_j,c^{\dagger}_l \right) & = & -i \int^{\infty}_0 {\rm d}t\, e^{i \omega t} \langle \mathrm{gs} | c^{\dagger}_l c^{\phantom{\dagger}}_j(t) | \mathrm{gs} \rangle\,,
\end{eqnarray}
which are related to the already introduced spectral functions 
\begin{eqnarray}
\langle \langle c^{\phantom{\dagger}}_j c^{\dagger}_l \rangle \rangle^{\mathrm{h}} _{\omega}=-\frac{1}{\pi} {\rm Im}\,G^{\mathrm{h}}\left(c^{\phantom{\dagger}}_j,c^{\dagger}_l \right)\,,\nonumber\\
\langle \langle c^{\dagger}_l c^{\phantom{\dagger}}_j \rangle \rangle^{\mathrm{e}} _{\omega}=-\frac{1}{\pi} {\rm Im}\,G^{\mathrm{e}}\left(c^{\phantom{\dagger}}_j,c^{\dagger}_l \right)\,.
\end{eqnarray}
Using the transformations \eqref{trans1} and \eqref{trans2} one may express $G^{\mathrm{e},\mathrm{h}}\left(c^{\phantom{\dagger}}_j,c^{\dagger}_l \right)$ as a linear combination of the Green's functions defined in terms of the Majorana fermions $G^{\mathrm{e},\mathrm{h}}\left(\Gamma_a,\Gamma_b \right)$. However, the only contributions to the in-gap spectral functions come from the zero-modes, i.e., from $a,b \in \{L,R\}$, and the corresponding functions can be obtained directly from the effective Hamiltonian \eqref{heff}, 
\begin{eqnarray}
G^{\mathrm{h}}\left(\Gamma_L, \Gamma_L \right)=G^{\mathrm{h}}\left(\Gamma_R, \Gamma_R \right)&=&\frac{1}{\omega-|\varepsilon|+i \eta}\,,\nonumber \\
G^{\mathrm{e}}\left(\Gamma_L, \Gamma_L \right)=G^{\mathrm{e}}\left(\Gamma_R, \Gamma_R \right)&=&\frac{1}{\omega+|\varepsilon|+i \eta}\,.
\label{greenoffset}
\end{eqnarray}
The Green's functions determine the in-gap peak in the left part of the system
\begin{equation}
G^{\alpha}\left(c^{\phantom{\dagger}}_j,c^{\dagger}_j \right) = \frac{V^2_{L,2j}+V^2_{L,2j-1}}{4}\,G^{\alpha}\left(\Gamma_L,\Gamma_L \right)\,,
\label{diaggreen}
\end{equation}
with $\alpha\in\{\mathrm{e},\mathrm{h}\}$, and a similar expression holds for the peak in its right side. Utilizing the orthogonality of $\hat V$, one may explicitly sum up the Green's functions over the lattice sites
\begin{equation}
\sum_j G^{\alpha}\left(c^{\phantom{\dagger}}_j,c^{\dagger}_j \right) = \frac{1}{4}G^{\alpha}\left(\Gamma_L,\Gamma_L \right)\,,
\label{sumrule}
\end{equation}
where the sum over $j$ contains few sites at the edge of the system due to the exponential decay of the $\hat V$ elements. The result Eq.~\eqref{sumrule} explains why the total spectral weights originating from $\sum_j G^{\alpha}$ equal $1/4$, while the total spectral weights of the peaks in LDOS equal $1/2$. Similar discussion of the nonlocal centrosymmetric spectral functions $\langle\langle c^{\phantom{\dagger}}_{\ell}\,c^{\dagger}_{L-\ell+1} \rangle\rangle_{\epsilon_{\mathrm{F}}}^{\mathrm{h}}$ and $\langle\langle c^{\dagger}_{\ell}\,c^{\phantom{\dagger}}_{L-\ell+1} \rangle\rangle_{\epsilon_{\mathrm{F}}}^{\mathrm{e}}$ can be found in the Supplementary Note~3.

%%%%%%%%%%%%%%%%%%%%%%%%%%%%%%%%%%%%%%%%%%%%%%%%%%%%%%%%%%%%%%%%%%%%%%%%%%%%%%%%%
\vspace{0.5em}
\noindent {\bf Von Neumann entanglement entropy.}
$S_{\mathrm{vN}}(\ell)$ measures entanglement between two subsystems containing, respectively, $\ell$ and $L-\ell$ sites, and can be easily calculated within DMRG via the reduced density matrix $\rho_\ell$, i.e., \mbox{$S_{\mathrm{vN}}(\ell)=-\mathrm{Tr}\rho_\ell\ln\rho_\ell$}. The results presented in Fig.~\ref{fig6} depict the system divided into two equal halves, $\ell=L/2$. The full spatial dependence of $S_{\mathrm{vN}}(\ell)$ is presented in the Supplementary Note~5.

%%%%%%%%%%%%%%%%%%%%%%%%%%%%%%%%%%%%%%%%%%%%%%%%%%%%%%%%%%%%%%%%%%%%%%%%%%%%%%%%%
\vspace{0.5em}
\noindent {\bf Superconducting amplitudes.}
The $s$-wave and $p$-wave SC can be detected with singlet $\Delta_{\mathrm{S}}$ and triplet $\Delta_{\mathrm{T}}$ amplitudes, respectively, defined as
\begin{eqnarray}
\Delta_{\mathrm{S}}&=&\sum_{\ell=L/4}^{3L/4}\left|\Delta_{\mathrm{S},\ell}\right|\,,\nonumber\\
\Delta_{\mathrm{T}}&=&\sum_{\ell=L/4}^{3L/4}\left(\left|\Delta_{\mathrm{T0},\ell}\right|+\left|\Delta_{\mathrm{T\downarrow},\ell}\right|+\left|\Delta_{\mathrm{T\downarrow},\ell}\right|
\right)\,,
\label{scamp}
\end{eqnarray}
with
\begin{eqnarray}
\Delta_{\mathrm{S},\ell}&=&\left\langle c^{\dagger}_{\ell,\uparrow}c^{\dagger}_{\ell+1,\downarrow} - c^{\dagger}_{\ell,\downarrow}c^{\dagger}_{\ell+1,\uparrow}\right\rangle\,,\nonumber\\
\Delta_{\mathrm{T0},\ell}&=&\left\langle c^{\dagger}_{\ell,\uparrow}c^{\dagger}_{\ell+1,\downarrow} + c^{\dagger}_{\ell,\downarrow}c^{\dagger}_{\ell+1,\uparrow}\right\rangle\,,\nonumber\\
\Delta_{\mathrm{T\uparrow},\ell}&=&\left\langle c^{\dagger}_{\ell,\uparrow}c^{\dagger}_{\ell+1,\uparrow}\right\rangle\,,\quad
\Delta_{\mathrm{T\downarrow},\ell}=\left\langle c^{\dagger}_{\ell,\downarrow}c^{\dagger}_{\ell+1,\downarrow}\right\rangle\,.
\end{eqnarray}

%================================================================================

%================================================================================
\vspace{1.0em}
\noindent {\bf Acknowledgments}
\noindent J.~Herbrych and M. {\'S}roda acknowledge support by the Polish National Agency for Academic Exchange (NAWA) under contract PPN/PPO/2018/1/00035 and, together with M. Mierzejewski, by the National Science Centre (NCN), Poland via project 2019/35/B/ST3/01207. The work of G. Alvarez was supported by the Scientific Discovery through Advanced Computing (SciDAC) program funded by the US DOE, Office of Science, Advanced Scientific Computer Research and Basic Energy Sciences, Division of Materials Science and Engineering. The development of the DMRG++ code by G.~Alvarez was conducted at the Center for Nanophase Materials Science, sponsored by the Scientific User Facilities Division, BES, DOE, under contract with UT-Battelle. E.~Dagotto was supported by the US Department of Energy (DOE), Office of Science, Basic Energy Sciences (BES), Materials Sciences and Engineering Division. A part of the calculations was carried out using resources provided by the Wroc{\l}aw Centre for Networking and Supercomputing.

%================================================================================
\vspace{1.0em}
\noindent {\bf Author contribution}
\noindent J.H., M.M., and E.D. planned the project. G.A. developed the \textsc{DMRG++} computer program. J.H. and M.\'{S}. performed the numerical simulations. J.H., M.\'{S}., M.M., and E.D. wrote the manuscript. All co-authors provided comments on the paper.

%================================================================================
\vspace{1.0em}
\noindent {\bf Additional information}
\noindent Supplementary Information accompanies this paper. Correspondence and requests for materials should be addressed to J.H. (email: \href{mailto:jacek.herbrych@pwr.edu.pl}{jacek.herbrych@pwr.edu.pl}).

%================================================================================
\vspace{1.0em}
\noindent {\bf Data availability}
\noindent The data that support the findings of this study are available from the corresponding author upon request.

%================================================================================
\vspace{1.0em}
\noindent {\bf Competing Interests}
\noindent The authors declare no competing interests.

%================================================================================
\clearpage
\appendix
\starttocentries
\setcounter{page}{1}
\setcounter{figure}{0}
\setcounter{equation}{0}
\newcommand{\rom}[1]{\uppercase\expandafter{\romannumeral #1\relax}}
\renewcommand{\refname}{Supplemental References}
\renewcommand{\figurename}{Supplementary Figure}
%\renewcommand{\thefigure}{S\arabic{figure}}
%\renewcommand{\citenumfont}[1]{S#1}
%\renewcommand{\bibnumfmt}[1]{[S#1]}
%\renewcommand{\thepage}{S\arabic{page}}
%\renewcommand{\theequation}{S\arabic{equation}}
%================================================================================
\onecolumngrid
\begin{center}
{\bf \uppercase{Supplementary Information}} for:\\
\vspace{10pt}
{\bf \large Interaction-induced topological phase transition and Majorana edge states\\in low-dimensional orbital-selective Mott insulators}\\
\vspace{10pt}
by J. Herbrych, M. {\'S}roda, G. Alvarez, M. Mierzejewski, and E. Dagotto
\end{center}
\vspace{10pt}
\tableofcontents
\vspace{10pt}

%================================================================================
\clearpage
\newpage
\section{Supplementary Note 1. Stability of the hybrid BdG-DMRG algorithm}

In the main text, we have presented how the spatial profiles of the pairing fields (PF) in the BCS system $\Delta^\mathrm{BCS}_{i}$ converge in subsequent steps of the iteration procedure. Here, as supplementary information, we will discuss the convergence of PF $\Delta^\mathrm{OSMP}_\ell$ obtained for the 1D OSMP system. Furthermore, we will test the stability of the introduced procedure and demonstrate that the obtained results are independent of the initial choice of the PF.

From the results presented in the main text, it is evident that the PF converge to almost uniform (spatially independent) values within the 1D OSMP system. Consequently, to simplify the presentation of the results, we will discuss only the behaviour of the average PF, $\sum_\ell \Delta^\mathrm{OSMP}_\ell/L$. In Supplementary~Figure~\ref{figS2} we present the iteration dependence of the latter for various initial starting points. As explained in the Methods section of the main text, our DMRG-BdG algorithm can be started from arbitrary amplitudes in the 1D OSMP system. We have considered: (1) zero PF (the result discussed in the main text), (2) constant PF $\Delta^\mathrm{OSMP}_\ell=0.1$ and $\Delta^\mathrm{OSMP}_\ell=1.0$, and (3) random PF drawn from a box distribution of widths $[0.0\,,0.1]$, $[0.0\,,1.0]$, $[-0.1\,,0.1]$, and $[-1.0\,,0.1]$. Several conclusions can be obtained directly from the results for $U/W=1$ (trivial phase, Supplementary~Figure~\ref{figS2}a) and $U/W=2$ (topological phase, Supplementary~Figure~\ref{figS2}b): (i) For all considered cases, the converged spatial profiles are almost uniform within the OSMP chain. This is best exemplified by the results presented in the right column of Fig.~2b of the main text, where we show the convergence of the hybrid procedure for the case when iterations are initialized by random PF, i.e., $\Delta^\mathrm{OSMP}_\ell\in[0,1]$. (ii) For all considered initial PF, the results converge to the same - interaction dependent - value. This nontrivial result shows that the hybrid BdG-DMRG procedure is numerically very stable. (iii) The quick convergence of the hybrid procedure holds true for the entire range of the Coulomb repulsion considered in the present studies: see results for $\sum_\ell \Delta^\mathrm{OSMP}_\ell/L$ shown in Supplementary~Figure~\ref{figS2}c and Supplementary~Figure~\ref{figS2}d.

%--------------------------------------------------------------------------------
\begin{figure}[!htb]
\includegraphics[width=1.0\textwidth]{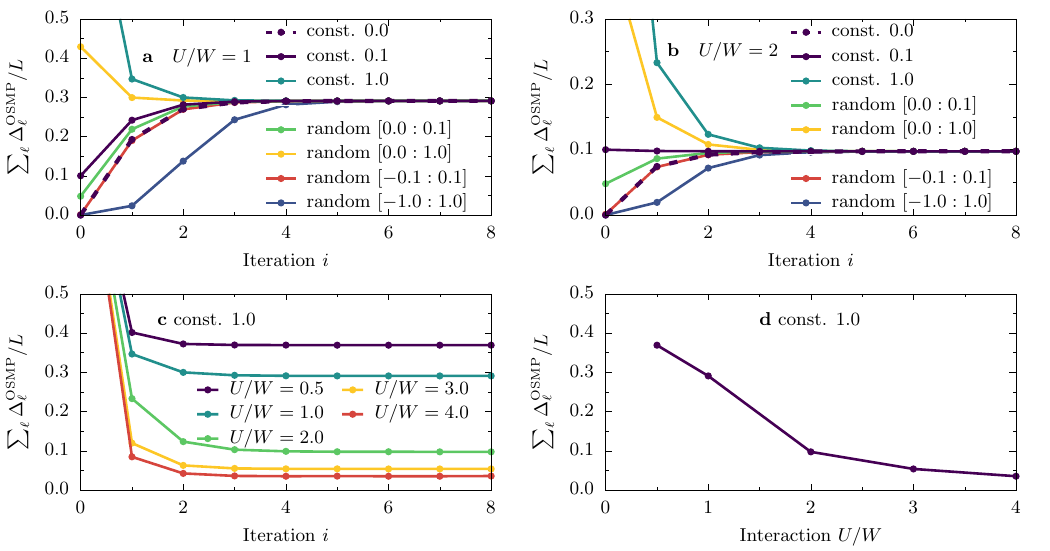}
\caption{{\bf Pairing fields within the 1D OSMP system.} {\bf a,b} Iteration convergence of the average pairing field amplitudes $\sum_\ell \Delta^\mathrm{OSMP}_\ell/L$ as obtained for {\bf a} $U/W=1$ and {\bf b} $U/W=2$, starting from various initial configuration of $\Delta^\mathrm{OSMP}_\ell$. Calculated for $L=36$, $\overline{n}=0.5$, and $V=4\,\mathrm{[eV]}$. All units in eV. See the Methods section of the main text for details. {\bf c} depicts the interaction $U$ dependence of the convergence for the case of an initial $\Delta^\mathrm{OSMP}_\ell=1\,\mathrm{[eV]}$. The asymptotic value of $\sum_\ell \Delta^\mathrm{OSMP}_\ell/L$ as function of $U$ is given in {\bf d}.}
\label{figS2}
\end{figure}
%--------------------------------------------------------------------------------

%================================================================================
\clearpage
\newpage
\section{Supplementary Note 2. Ladder geometry considerations}

In the main text, we have shown that the generalized Kondo-Heisenberg (gKH) model on the chain geometry can support Majorana zero-energy modes (MZM) when in the presence of a superconducting (SC) pairing field $\Delta_{\mathrm{SC}}$. Here, we will show that the key ingredients necessary to support the MZM in the gKH model are also present in the ladder geometry, i.e. the block-spiral magnetic state (see Supplementary~Figure~\ref{figS1}a for a sketch) and single-particle spectra with parity-breaking quasi-particles. We will consider a spatially isotropic ladder with $t_{\parallel}=t_{\perp}\equiv t_{\mathrm{i}}$, the latter hopping defined in the main text, and choose filling $\overline{n}=1.75$, which supports block-magnetism at $U\sim W$ \cite{SHerbrych2018,SHerbrych2019-1}. Although accurate calculations within the grand-canonical ensemble (needed with finite pairing field $\Delta_{\mathrm{SC}}\ne0$) are numerically too demanding (due to the doubling of the lattice sites on the ladder of $L$ rungs), precise canonical calculations of the static structure factor can still be performed, i.e., 
\begin{equation}
S(q_\parallel,q_\perp)=\sum_{\ell,\ell^\prime}\sum_{r,r^\prime}
\mathrm{e}^{+iq_\perp(r-r^\prime)}\mathrm{e}^{+iq_\parallel(\ell-\ell^\prime)}
\langle \mathbf{T}_{\ell,r}\cdot \mathbf{T}_{\ell^\prime,r^\prime}\rangle\,,
\end{equation}
where $\mathbf{T}_{\ell,r}=\mathbf{S}_{\ell,r}+\mathbf{s}_{\ell,r}$, and $(\ell,r)$ represent the leg and rung number, respectively. Our results in Supplementary~Figure~\ref{figS1}b reveal that the $S(q_\parallel,q_\perp)$ lies at incommensurate values of the wavevectors, the one of the block-spiral magnetic state signatures~\cite{SHerbrych2019-2}. Another feature of the latter is the existence of two cosine-like bands in the single-particle spectral (see also the discussion in the next section) $A(q_{\parallel},q_{\perp},\omega)=A^{\mathrm{e}}(q_{\parallel},q_{\perp},\omega)+A^{\mathrm{h}}(q_{\parallel},q_{\perp},\omega)$ near the Fermi level $\omega\sim\epsilon_{\mathrm{F}}$, where
\begin{eqnarray}
A^{\mathrm{e}}(q_{\parallel},q_{\perp},,\omega)&=&\sum_{\ell}\sum_{r,r^\prime} \mathrm{e}^{+iq_\perp(r-r^\prime)}\mathrm{e}^{-iq_{\parallel}(\ell-L/2)} \,\langle\langle c^{\phantom{\dagger}}_{\ell,r}\,c^{\dagger}_{L/2,r^\prime} \rangle\rangle_{\omega}^{\mathrm{e}}\,,\nonumber\\
A^{\mathrm{h}}(q_{\parallel},q_{\perp},,\omega)&=&\sum_{\ell}\sum_{r,r^\prime} \mathrm{e}^{+iq_\perp(r-r^\prime)}\mathrm{e}^{+iq_{\parallel}(\ell-L/2)} \,\langle\langle c^{\dagger}_{\ell,r}\,c^{\phantom{\dagger}}_{L/2,r^\prime} \rangle\rangle_{\omega}^{\mathrm{h}}\,.
\end{eqnarray}
The results presented in Supplementary~Figure~\ref{figS1}c (for $L=36$ rungs, $U/W=2.2$, $J_{\mathrm{H}}/U=0.25$, and $\overline{n}=1.75$) are consistent with this scenario and resemble the chain geometry results. Consequently, it is reasonable to assume that the influence of a finite pairing field $\Delta_{\mathrm{SC}}\ne0$ will lead to a topological superconducting state and the emergence of the MZM also on the ladder lattice. 

%--------------------------------------------------------------------------------
\begin{figure}[!htb]
\includegraphics[width=0.5\textwidth]{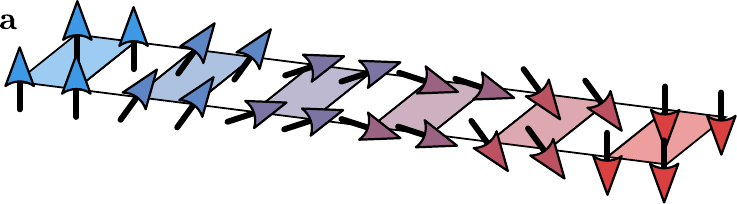}\\
\includegraphics[width=0.9\textwidth]{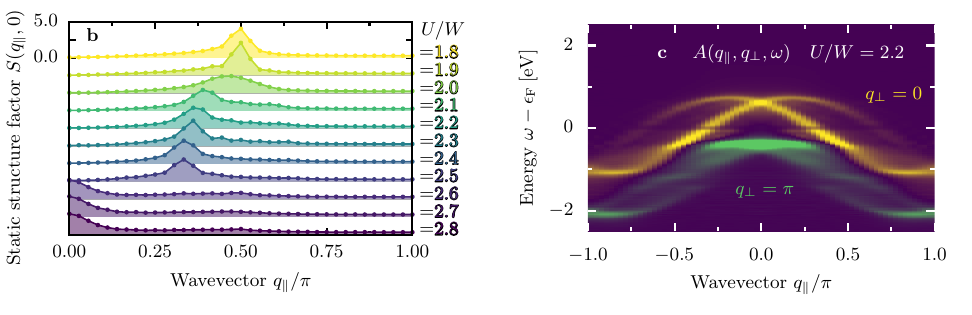}
\caption{{\bf Block-spiral state on the ladder geometry.} {\bf a} Sketch of the block-spiral state on the ladder geometry. Color (shaded) area represents rigidly rotating $2\times2$ FM blocks. {\bf b} Interaction dependence of the symmetric component $q_{\perp}=0$ of the static structure factor $S(q_{\parallel},q_{\perp}=0)$. {\bf c} Wavevector dependence of both components (symmetric and antisymmetric) of the single-particle spectral function \mbox{$A(q_{\parallel},q_{\perp}=0,\omega)+A(q_{\parallel},q_{\perp}=\pi,\omega)$} near the Fermi level in the block-spiral phase ($\eta=2\delta\omega$ and $\delta\omega=0.02$~[eV]). All results calculated for the gKH ladder of $L=36$ rungs, $J_\mathrm{H}/U=0.25$, $\overline{n}=1.75$, and $\Delta_{\mathrm{SC}}=0$.}
\label{figS1}
\end{figure}
%--------------------------------------------------------------------------------

%================================================================================
\clearpage
\newpage
\section{Supplementary Note 3. Spectral functions}

In Fig.~4 of the main text, we have shown the spatial dependence of the local density-of-states (LDOS) at the Fermi level, together with equal contributions of the electron and hole components, as expected for MZM. Here, in Supplementary~Figure~\ref{figS3}, we show that the same holds in frequency $\omega$ space. Furthermore, in the same figure we show that both spin components contribute equally (within our numerical precision).

%--------------------------------------------------------------------------------
\begin{figure}[!htb]
\includegraphics[width=1.0\textwidth]{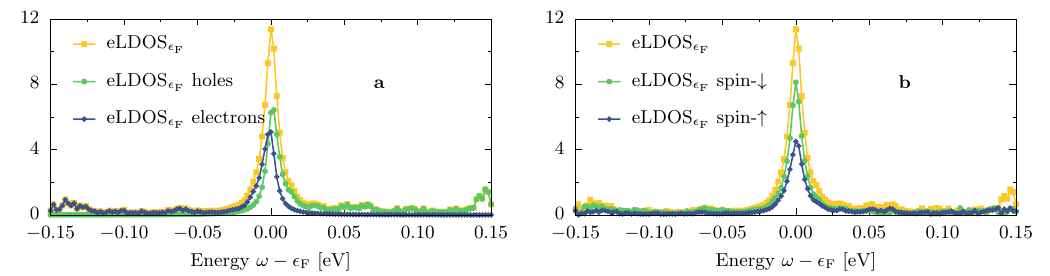}
\caption{{\bf Components of the Majorana edge states.} Frequency $\omega$ dependence of the edge-LDOS ($\ell=1$) near the Fermi level $\omega\sim\epsilon_{\mathrm{F}}$ Panel {\bf a} depicts electron and hole contributions, while panel {\bf b} the $\downarrow$- and $\uparrow$-spin component. Calculated for $L=36$, $U/W=2$, $\Delta_{\mathrm{SC}}/W\simeq0.5$, and $\overline{n}=0.5$.}
\label{figS3}
\end{figure}
%--------------------------------------------------------------------------------

The same reasoning used for the LDOS in the main text can also be applied to the off-diagonal functions $G^{\alpha}\left(c^{\phantom{\dagger}}_j,c^{\dagger}_l \right)$ where sites $j$ and $l$ belong to the left $j<L/2$ and the right $l>L/2$ portions of the system. Then, it can be shown that
\begin{eqnarray}
G^{\alpha}\left(c^{\phantom{\dagger}}_j,c^{\dagger}_l \right)
&=&\frac{V_{L,2j-1}V_{R,2l-1}+V_{L,2j}V_{R,2l}}{4}\,G^{\alpha}\left(\Gamma_L,\Gamma_R \right), \nonumber \\
&+i& \frac{V_{L,2j}V_{R,2l-1}-V_{L,2j-1}V_{R,2l}}{4}\,G^{\alpha}\left(\Gamma_L,\Gamma_R \right)\,,
\label{goff}
\end{eqnarray}
with 
\begin{eqnarray}
G^{\mathrm{h}}\left(\Gamma_L, \Gamma_R \right)=-G^{\mathrm{h}}\left(\Gamma_R, \Gamma_L \right)&=&\frac{i \, {\rm sgn}(\varepsilon)}{\omega-|\varepsilon|+i \eta}\,, \nonumber \\
G^{\mathrm{e}}\left(\Gamma_L, \Gamma_R \right)=-G^{\mathrm{e}}\left(\Gamma_R, \Gamma_L \right)&=&\frac{- i \, {\rm sgn}(\varepsilon)}{\omega+|\varepsilon|+i \eta}\,.
\label{gg}
\end{eqnarray}
Since the considered Hamiltonian is real, the spectral functions $\langle \langle c^{\phantom{\dagger}}_j c^{\dagger}_l \rangle \rangle^{\mathrm{h}}_{\omega}$ and $\langle \langle c^{\dagger}_j c^{\phantom{\dagger}}_l \rangle \rangle^{\mathrm{e}}_{\omega}$ should be real as well. Given that the weights of $\langle \langle \Gamma_L \Gamma_R \rangle \rangle^{\alpha}_{\omega}$ are purely imaginary [see Supplementary~Eq.~\eqref{gg}], the upper line in Supplementary~Eq.~\eqref{goff} should vanish. Indeed, for real Hamiltonians, $\Gamma_L$ (and also $\Gamma_R$) contains $\gamma_j$ with only even or odd $j$. In other words, $\Gamma_L$ contains only $\gamma_{2j}$ and $\Gamma_R$ contains only $\gamma_{2j-1}$ or {\it vice versa}. Without losing generality, we may choose the former possibility,
\begin{eqnarray}
G^{\alpha}\left(c^{\phantom{\dagger}}_j,c^{\dagger}_l \right)&=&i \frac{V_{L,2j}V_{R,2l-1}}{4}\,G^{\alpha}\left(\Gamma_L,\Gamma_R \right) \nonumber \\
 G^{\alpha}\left(c^{\phantom{\dagger}}_l,c^{\dagger}_j \right)&=&-i \frac{V_{R,2l-1}V_{L,2j}}{4}\,G^{\alpha}\left(\Gamma_R,\Gamma_L \right) = G^{\alpha}\left(c^{\phantom{\dagger}}_j,c^{\dagger}_l \right)\,,
\end{eqnarray}
and obtain the spectral functions shown in Supplementary~Figure~\ref{figS4} and Fig.~4b of the main text
\begin{eqnarray}
\langle \langle c^{\phantom{\dagger}}_l c^{\dagger}_{L-l+1} \rangle \rangle^{\mathrm{h}}_{\omega}&=& -\frac{1}{4}\left\{
{V_{L,2l}V_{R,2L-2l+1}\,\,\text{for}\,\,l<L/2
\atop
V_{L,2L-2l+2}V_{R,2l-1}\,\,\text{for}\,\,l>L/2}
\right\}\,{\rm sgn}(\varepsilon)\,\delta(\omega-|\varepsilon|)\,,\nonumber \\
\nonumber \\
\langle \langle c^{\dagger}_{l} c^{\phantom{\dagger}}_{L-l+1}\rangle \rangle^\mathrm{e}_{\omega}&=&+\frac{1}{4}\left\{
V_{L,2l}V_{R,2L-2l+1}\,\,\text{for}\,\,l<L/2
\atop
V_{L,2L-2\ell+2}V_{R,2l-1}\,\,\text{for}\,\,l>L/2
\right\}\,{\rm sgn}(\varepsilon)\,\delta(\omega+|\varepsilon|)\,,
\label{ofdiaggreen}
\end{eqnarray}
where the electron and hole contributions arise with opposite signs, as it is also visible in Supplementary~Figure~\ref{figS4} and Fig.~4b of the main text. Finally, it is reasonable to assume that the spatial profiles of the MZMs at both system edges are mutually symmetric, i.e., $|V_{R,2L-2l+1}|\simeq|V_{L,2l}|$. Then, comparing Eq.~15 of the main text and Supplementary~Eq.~\eqref{ofdiaggreen} we obtain a mirroring of the diagonal (local) and off-diagonal spectral functions
\begin{eqnarray}
|\langle \langle c^{\phantom{\dagger}}_l c^{\dagger}_{L-l+1} \rangle \rangle^{\mathrm{h}}_{\omega}|& \simeq &|\langle \langle c^{\phantom{\dagger}}_l c^{\dagger}_{l} \rangle \rangle^{\mathrm{h}}_{\omega}|, \nonumber \\ 
|\langle \langle c^{\dagger}_{L-l+1} c^{\phantom{\dagger}}_l \rangle \rangle^{\mathrm{e}}_{\omega}|& \simeq &|\langle \langle c^{\dagger}_{l} c^{\phantom{\dagger}}_l \rangle \rangle^{\mathrm{e}}_{\omega}|\,, 
\end{eqnarray}
which is reasonably well reproduced by the numerical results. 
 
%--------------------------------------------------------------------------------
\begin{figure}[!htb]
\includegraphics[width=0.6\textwidth]{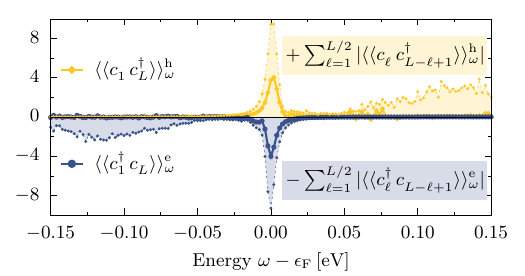}
\caption{{\bf Off-diagonal spectral functions.} Frequency dependence of centrosymmetric spectral functions, Supplementary~Eq.~\eqref{ofdiaggreen}, at the edge of the system $\ell=1$ (solid points). We also present spatially integrated, according to Eq.~16 of the main text, spectral functions as colored area. Results shown were calculated for $L=36$, $U/W=2$, $\Delta_{\mathrm{SC}}/W\simeq0.5$, and $\overline{n}=0.5$.}
\label{figS4}
\end{figure}
%--------------------------------------------------------------------------------

%================================================================================
\clearpage
\newpage
\section{Supplementary Note 4. Parameter dependence}

In this section, we will discuss the pairing field $\Delta_{\mathrm{SC}}$ dependence of our results. Let us first focus on the single-particle spectral function $A(q,\omega)$ [see Eq.~7 of the main text]. In Supplementary~Figure~\ref{figS5}a and Supplementary~Figure~\ref{figS5}b we show $A(q,\omega)$ for systems (at electronic filling $\overline{n}=0.5$) without pairing field $\Delta_{\mathrm{SC}}=0$ for two representative values of the interaction: $U/W=1$ and $U/W=2$, i.e., in the block-collinear and block-spiral magnetic phases. Both spectra exhibit a finite density-of-states (DOS) at the Fermi level $\epsilon_{\mathrm{F}}$. In the case of the block-spiral phase at $U/W=2$, Supplementary~Figure~\ref{figS5}b, one can observe two bands of quasiparticles: left and right movers reflecting the two possible rotations of the spirals. It is obvious from these results that the quasiparticles break the parity symmetry; i.e., going from $q\to-q$ momentum changes the quasiparticle character, as expected for a spiral state. It is also worth noting that for the block-magnetic order ($U/W=1$ and $\Delta_{\mathrm{SC}}=0$) one can observe \cite{SHerbrych2019-2} the V-like shape of DOS in the vicinity of $\epsilon_{\mathrm{F}}$. The latter indicates a semiconductor-like behaviour which was also experimentally found \cite{SLei2011} in the $2\times2$ block-magnetic ladder compound BaFe$_2$Se$_3$. This result shows our model's strength and relevance for realistic investigations of the iron-based materials from the 123 family.

%--------------------------------------------------------------------------------
\begin{figure}[!htb]
\includegraphics[width=0.8\textwidth]{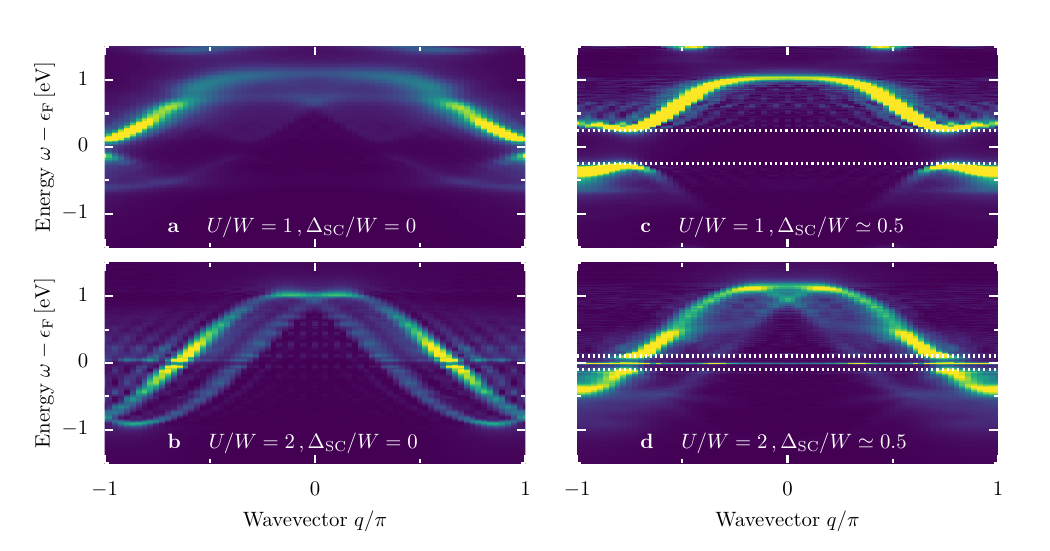}
\caption{{\bf Single-particle spectra for block-collinear and block-spiral magnetism.} Single-particle spectra $A(q,\omega)$ of the gKH model using $L=36$ sites and electronic filling $\overline{n}=0.5$ for {\bf a} $U/W=1\,,\Delta_{\mathrm{SC}}=0$, {\bf b} $U/W=2\,,\Delta_{\mathrm{SC}}=0$, {\bf c}, $U/W=1\,,\Delta_{\mathrm{SC}}/W\simeq 0.5$, and {\bf d} $U/W=2\,,\Delta_{\mathrm{SC}}/W\simeq 0.5$.}
\label{figS5}
\end{figure}
%--------------------------------------------------------------------------------

The pairing field $\Delta_{\mathrm{SC}}\ne0$ has a striking effect on these two phases, see Supplementary~Figure~\ref{figS5}c and Supplementary~Figure~\ref{figS5}d where we present results for $\Delta_{\mathrm{SC}}/W\simeq 0.5$. As discussed in detail in the main text, the pairing field leads to the appearance of MZM in the spiral phase ($U/W=2$, Supplementary~Figure~\ref{figS5}d), with the flat $\delta$-mode inside the superconducting gap. On the other hand, for the collinear block-magnetic phase ($U/W=1$, Supplementary~Figure~\ref{figS5}c) we observe only a trivial opening of the SC gap, without any in-gap states. These results indicate that the Hubbard interaction strength $U$, as the main driver between the two magnetic states, plays a crucial role in the stabilization of the MZM.

%--------------------------------------------------------------------------------
\begin{figure}[!htb]
\includegraphics[width=1.0\textwidth]{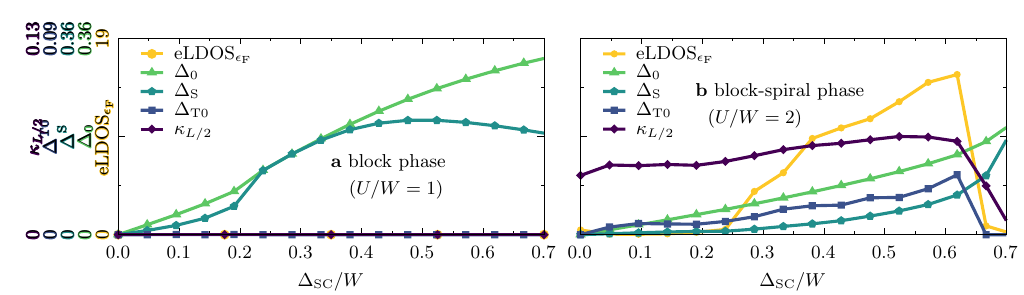}
\caption{{\bf Phase diagram.} Pairing field $\Delta_{\mathrm{SC}}$ dependence of: (i) the value of edge-LDOS at the Fermi level $\epsilon_{\mathrm{F}}$ (eLDOS), (ii) chirality correlation function $\langle \boldsymbol{\kappa}_{L/4}\cdot \boldsymbol{\kappa}_{3L/4} \rangle$ at $L/2$ distance ($\boldsymbol{\kappa}_{L/2}$), (iii) amplitudes of local and non-local SC singlet amplitudes, $\Delta_{0}$ and $\Delta_{\mathrm{S}}$, respectively, together with triplet component $\Delta_{\mathrm{T0}}$. Panel {\bf a} shows results for the block-collinear magnetic phase ($U/W=1$), while panel {\bf b} for the block-spiral phase ($U/W=2$). All results were calculated for $L=36$ and $\overline{n}=0.5$.}
\label{figS6}
\end{figure}
%--------------------------------------------------------------------------------

In order to investigate all these aspects in more detail, in Supplementary~Figure~\ref{figS6} we present the pairing field $\Delta_{\mathrm{SC}}$ dependence of the quantities discussed in the main text, i.e.: (i) value of edge-LDOS at the Fermi level $\epsilon_{\mathrm{F}}$ (eLDOS), (ii) chirality correlation function $\langle \boldsymbol{\kappa}_{L/4}\cdot \boldsymbol{\kappa}_{3L/4} \rangle$ at $L/2$ distance ($\boldsymbol{\kappa}_{L/2}$), and (iii) amplitudes of extended (non-local) SC singlet and triplet amplitudes, $\Delta_{\mathrm{S}}$ and $\Delta_{\mathrm{T0}}$, respectively [see Eq.~17 of the main text]. Furthermore, in the same figure we present the value of the on-site pairing amplitude
\begin{equation}
\Delta_0=\frac{2}{L}\sum_{\ell=L/4}^{3L/4} \big|c^{\dagger}_{\ell,\uparrow}c^{\dagger}_{\ell,\downarrow}\big|\,.
\end{equation}

For the collinear block-magnetic phase ($U/W=1$, Supplementary~Figure~\ref{figS6}a) we observe that the $\Delta_{\mathrm{SC}}$ does not induce any topological phase transitions. For all considered values of the pairing field, $0<\Delta_{\mathrm{SC}}/W\lesssim 0.7$, the eLDOS, $\boldsymbol{\kappa}_{L/2}$, and $\Delta_{\mathrm{T}}$ are zero. Only the singlet SC amplitudes, local $\Delta_0$ and non-local $\Delta_{\mathrm{S}}$, take a finite value. The behaviour of the $U/W=2$ case is strikingly different (see Supplementary~Figure~\ref{figS6}b). The chirality correlation function $\boldsymbol{\kappa}_{L/2}$ has a finite value already at $\Delta_{\mathrm{SC}}=0$, reflecting the block-spiral ordering at this interaction strength, and weakly changes till $\Delta_{\mathrm{SC}}/W\sim0.6$, after which it decays to zero. Simultaneously, the triplet SC amplitude $\Delta_{\mathrm{T0}}$ increases smoothly with the pairing-field $\Delta_{\mathrm{SC}}$ (together with singlet components $\Delta_0$ and $\Delta_{\mathrm{S}}$). It is worth noting that this behaviour is strikingly different from the $U$ variation presented in the main text, where we observed a sudden appearance of $\boldsymbol{\kappa}_{L/2}$ and $\Delta_{\mathrm{T0}}$ at a specific value of $U_{\mathrm{c}}/W=1.51$. Moreover, we remark again that the pairing field influences on the characteristics of the spiral, optimizing its shape from block to canonical, to better host the MZM.

The behaviour of the edge-LDOS in the spiral phase needs special attention. As evident from the results presented in Supplementary~Figure~\ref{figS6}b, the value of the latter becomes finite for $\Delta_{\mathrm{SC}}/W\gtrsim0.25$ and vanishes for $\Delta_{\mathrm{SC}}/W\sim0.7$ (together with the already discussed $\boldsymbol{\kappa}_{L/2}$ and $\Delta_{\mathrm{T0}}$). In order to explain the missing weight of edge LDOS for $\Delta_{\mathrm{SC}}/W\lesssim0.25$ let us investigate the frequency dependence of the hole $\langle\langle c^{\phantom{\dagger}}_1 c^{\dagger}_1 \rangle\rangle^{\mathrm{h}}_\omega$ and electron $\langle\langle c^{\dagger}_1 c^{\phantom{\dagger}}_1 \rangle\rangle^{\mathrm{e}}_\omega$ contributions to the edge-LDOS. As it is evident from the results in Supplementary~Figure~\ref{figS7}a, upon increasing $\Delta_{\mathrm{SC}}$ the peaks in the electron- and hole-like spectral functions approach each other (Supplementary~Figure~\ref{figS7}b shows in more detail the positions of both maxima). Within the accessible frequency resolution, both peaks are easily distinguishable for $\Delta_{\mathrm{SC}}/W\simeq 0.25$, while they merge into a single peak for $\Delta_{\mathrm{SC}}/W>0.3$.

%--------------------------------------------------------------------------------
\begin{figure}[!htb]
\includegraphics[width=1.0\textwidth]{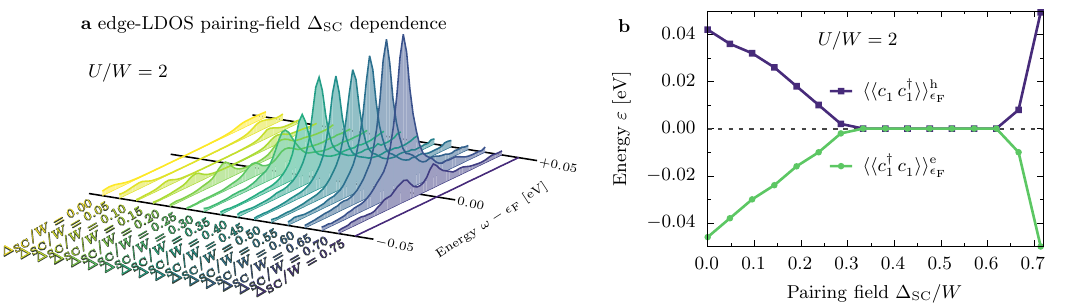}
\caption{{\bf Pairing field dependence of the edge local density-of-states.} {\bf a} Frequency $\omega$ dependence of edge ($\ell=1$) local density-of-states (LDOS) as a function of the pairing field $\Delta_{\mathrm{SC}}/W\simeq\{0.0\,,0.05\,,\dots,0.7\}$, as calculated for $L=36$, $\overline{n}=0.5$, and $U/W=2$. {\bf b} Pairing field dependence of the maximum position [i.e., offset energy $\varepsilon$, see Eq.~14 of the main text] of the hole $\langle\langle c^{\phantom{\dagger}}_1 c^{\dagger}_1 \rangle\rangle^{\mathrm{h}}_\omega$ and electron $\langle\langle c^{\dagger}_1 c^{\phantom{\dagger}}_1 \rangle\rangle^{\mathrm{e}}_\omega$ contributions to edge-LDOS, see Eq.~8 of the main text (based on the data presented in panel {\bf a}).}
\label{figS7}
\end{figure}
%--------------------------------------------------------------------------------

The behaviour described above is characteristic of systems hosting the MZM, i.e., despite the Majorana modes being located at the opposite edges of the studied chain, they overlap in any finite-$L$ system \cite{SStanescu2013,SRainis2013}. Due to this overlap, the peaks arise at a nonzero frequency $\omega=\pm \varepsilon$, see Eq.~14 of the main text. The clear splitting in the former case allows for a systematic finite-size study, shown in Supplementary~Figure~\ref{figS8}. In order to obtain well merged Majorana modes for $\Delta_{\mathrm{SC}}/W\simeq 0.25$, one needs to consider chains with at least $L\sim 60$ sites (see Supplementary~Figure~\ref{figS8}c), whereas systems half that size are sufficient for the case that is primarily studied in the present work, i.e., for $\Delta_{\mathrm{SC}}/W\simeq 0.5$ (see Supplementary~Figure~\ref{figS8}d). Furthermore, it is worth noting that if the chain is too short, then the remnants of the peaks become visible also in the middle of the system, as it is visible from results for $L=24$ in Supplementary~Figure~\ref{figS8}a and Supplementary~Figure~\ref{figS8}c. On the other hand, a clear single peak at the system's edge always coincides with a well-developed gap in the bulk, as shown in Supplementary~Figure~\ref{figS8}b and Supplementary~Figure~\ref{figS8}d. All these results consistently support the scenario that the nonzero splitting $\varepsilon$ originates from the overlap of the edge modes. In Supplementary~Figure~\ref{figS8}e we explicitly show that $\varepsilon$ decays exponentially with $L$, as expected for systems hosting the MZM \cite{SStanescu2013,SRainis2013}.
 
%--------------------------------------------------------------------------------
\begin{figure}[!htb]
\includegraphics[width=0.8\textwidth]{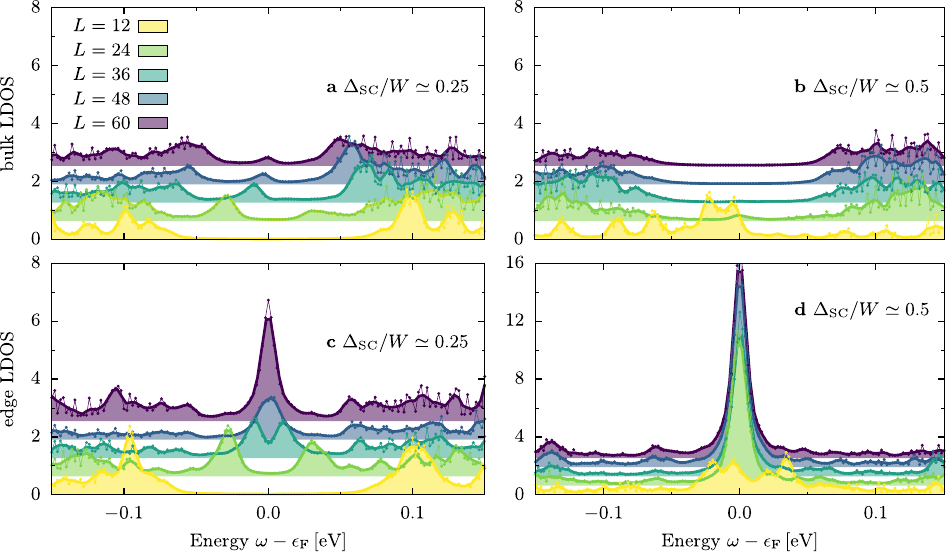}
\includegraphics[width=0.5\textwidth]{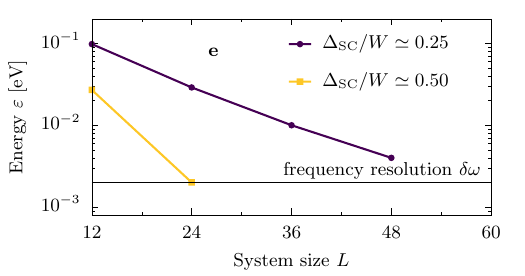}
\caption{{\bf Finite-size analysis.} System lengths $L=\{12\,,24\dots,60\}$ dependence of the local density-of-states {\bf a-b} in the middle of the chain representing the bulk ($\ell=L/2$) and {\bf c,d} at the edge ($\ell=1$) of the system, as calculated for $U/W=2$, {\bf a,c} $\Delta_{\mathrm{SC}}/W\simeq0.25$ and {\bf b,d} $\Delta_{\mathrm{SC}}/W\simeq 0.5$. {\bf e} System size $L$ dependence of the offset energy $\varepsilon$ [see Eq.~14 of the main text] for $\Delta_{\mathrm{SC}}/W\simeq\{0.25,0.5\}$ (based on the data presented in panels {\bf c} and {\bf d}).}
\label{figS8}
\end{figure}
%--------------------------------------------------------------------------------

Finally, we discuss the robustness of our results to modifications in the localized orbital interaction strength $U_K$. Within our model, the latter manifests as a change in the spin exchange integral $K=4t^2_{\mathrm{l}}/U_K$. Our results, presented in Supplementary~Figure~\ref{figS9}a, indicate that when the system is in the trivial phase, $U/W=1$ and $\Delta_\mathrm{SC}/W\simeq0.5$, only a singlet SC amplitude is present for all considered values of $U_K/W\in[0.2,3.0]$. For the topological phase at $U/W=2$, the results (the presence of the triplet SC amplitude) do not depend on $U_K$ as long as the spiral nature of magnetism is not destroyed, i.e., for $U_K/W\gtrsim0.8$. On the other hand, when the spin exchange integral $K\propto1/U_K$ dominates as an energy scale, the AFM ordering of the spins becomes energetically favorable [see Supplementary~Figure~\ref{figS9}b for the analysis of static structure factor $S(q)$] and, consequently, the system goes away from the topological phase. These results highlight the importance of the competing energy scales present in the multi-orbital OSMP system.

%--------------------------------------------------------------------------------
\begin{figure}[!htb]
\includegraphics[width=1.0\textwidth]{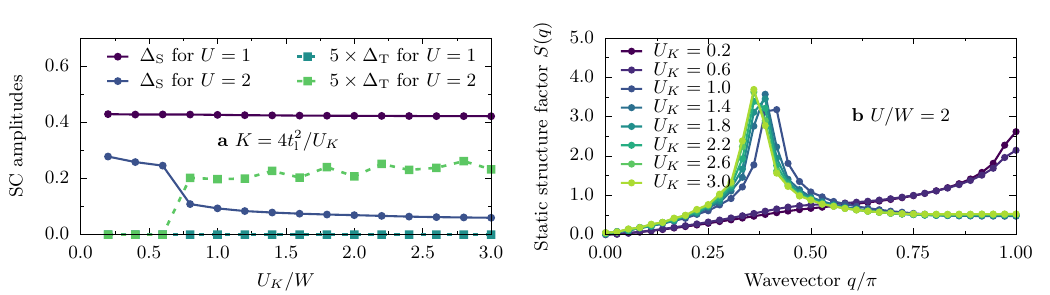}
\caption{{\bf Spin exchange analysis.} {\bf a} Dependence of the singlet $\Delta_\mathrm{S}$ and triplet $\Delta_\mathrm{T}$ SC amplitudes on the localized orbital interaction strength $U_K$, calculated for $L=36$, $U/W=1,2$, $\overline{n}=0.5$, and $\Delta_\mathrm{SC}/W\simeq0.5$. {\bf b} Static spin structure factor $S(q)$ dependence on the localized orbital interaction strength $U_K$, calculated for $L=36$, $U/W=2$, $\overline{n}=0.5$, and $\Delta_\mathrm{SC}/W\simeq0.5$.}
\label{figS9}
\end{figure}
%--------------------------------------------------------------------------------

%================================================================================
\clearpage
\newpage
\section{Supplementary Note 5. Entropy and dimer order}

\begin{figure}[!htb]
\includegraphics[width=0.8\textwidth]{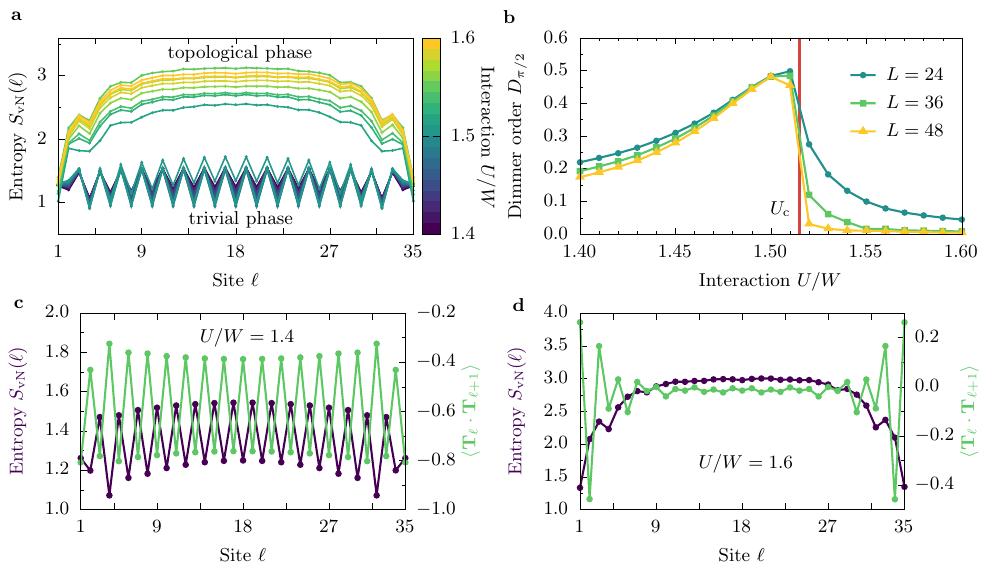}
\caption{{\bf Entropy $S_{\mathrm{vN}}$ and dimer order $D_{\pi/2}$.} {\bf a} Interaction \mbox{$U\in\{1.41,1.42,\dots,1.59,1.60\}$} dependence of the von Neumann entanglement entropy $S_{\mathrm{vN}}(\ell)$ of the subsystem of size $\ell$. Calculated for $L=36$, $\overline{n}=0.5$, and $\Delta_{\mathrm{SC}}/W \simeq 0.5$. {\bf b} Interaction dependence of the dimer order parameter $D_{\pi/2}$ as calculated for $L=24,36,38$, $\overline{n}=0.5$, and $\Delta_\mathrm{SC}/W\simeq0.5$. {\bf c-d} Site $\ell$ dependence of von Neumann entanglement entropy $S_{\mathrm{vN}}$ and local spin-spin correlation function $\langle \mathbf{T}_{\ell} \cdot \mathbf{T}_{\ell+1}\rangle$ (where $\mathbf{T}_\ell=\mathbf{S}_\ell+\boldsymbol{s}_\ell$ is the total on-site spin) {\bf c} below ($U/W=1.4$) and {\bf d} above ($U/W=1.6$) the topological phase transition. Results calculated for $L=36$, $\Delta_{\mathrm{SC}}/W\simeq0.5$, and $\overline{n}=0.5$.}
\label{figS10}
\end{figure}

In this section, we demonstrate that the interaction-induced topological phase transition at $U_c$ may be identified via studying the entanglement entropy. Supplementary~Figure~\ref{figS10}a shows the dependence of the von Neumann entropy $S_\mathrm{vN}(\ell)$ on the subsystem size, ${\ell}\le L$, in the vicinity of the transition, i.e., for $1.4<U/W<1.6$. Two characteristic behaviours emerge: for $U<U_{\mathrm{c}}\simeq1.51W$ $S_{\mathrm{vN}}(\ell)$ displays an oscillatory behaviour, while for $U>U_{\mathrm{c}}$ the entropy increases abruptly and becomes a smooth function of $\ell$. This sudden change in the entropy behaviour signifies the interaction-induced topological phase transition and the appearance of MZM.

We also argue that the topological transition in the OSMP chain is accompanied by a rapid change of the dimer order $D_{\pi/2}$ (see Supplementary~Figure~\ref{figS10}b). In Supplementary~Figures~\ref{figS10}c and \ref{figS10}d we have shown the entanglement entropy, respectively for the trivial and nontrivial phases, where $S_{\rm vN}(\ell)$ displays clear oscillations in the former case. To explain the physical origin of such oscillations we have also plotted the static spin-spin correlation function $\langle \mathbf{T}_{\ell}\cdot \mathbf{T}_{\ell+1} \rangle $. We can observe that the maxima of $|\langle \mathbf{T}_{\ell}\cdot \mathbf{T}_{\ell+1} \rangle| $ and $S_{\rm vN}(\ell)$ coincide. Recall to calculate the entanglement entropy, we split the system into two subsystems cutting the bond between sites $\ell$ and $\ell+1$. Whenever a bond with a large spin-spin correlation is cut, also the entanglement entropy is large. Therefore, we expect that the oscillatory behaviour of $S_{\rm vN}(\ell)$ is a direct manifestation of the dimer order. Indeed, in the topological phase the spin dimerization persists only at the very edges of the system, as shown in Supplementary~Figure~\ref{figS10}d, so that the (bulk) dimer order vanishes presumably as $1/L$, see Supplementary~Fig.~\ref{figS10}b. Due to the absence of bulk dimer order, the entanglement entropy smoothly changes with $\ell$, as demonstrated also in Supplementary~Figure~\ref{figS10}a.

%--------------------------------------------------------------------------------
%--------------------------------------------------------------------------------

Furthermore, the vanishing of the dimer order can also be observed in the behaviour of the chirality correlation function $\langle \boldsymbol{\kappa}_{L/2}\cdot\boldsymbol{\kappa}_{\ell}\rangle$ shown in Supplementary~Figure~\ref{figS11}. As expected in the $\Delta_{\mathrm{SC}}\to 0$ limit (for which $D_{\pi/2}\ne0$), the $\langle \boldsymbol{\kappa}_{L/2}\cdot\boldsymbol{\kappa}_{\ell}\rangle$ correlation displays a clear zig-zag–like pattern, reflecting the $\pi/2$-block nature of the spiral (for details see Ref.~\cite{SHerbrych2019-2}). It is evident from the presented results that the decay length of the $\boldsymbol{\kappa}$-correlation is not affected by the pairing field strength. On the other hand, the spatial details of $\langle \boldsymbol{\kappa}_{L/2}\cdot\boldsymbol{\kappa}_{\ell}\rangle$ change from a zig-zag to a smooth function of distance. The latter is consistent with the $D_{\pi/2}\to 0$ result in this region.

%--------------------------------------------------------------------------------
\begin{figure}[!htb]
\includegraphics[width=0.5\textwidth]{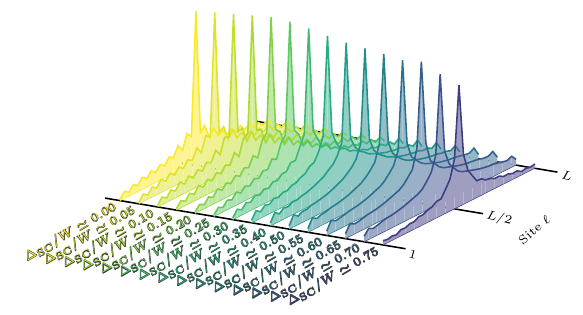}
\caption{{\bf Pairing field dependence of the spiral order.} Site dependence of the chirality correlation function $\langle \boldsymbol{\kappa}_{L/2} \cdot \boldsymbol{\kappa}_{\ell}\rangle$ as calculated for various strengths of the SC pairing fields $\Delta_\mathrm{SC}/W\simeq(0.00,0.05,\dots,0.75)$ ($\overline{n}=0.5$, $U/W=2$, $L=36$).}
\label{figS11}
\end{figure}
%--------------------------------------------------------------------------------

%================================================================================

%================================================================================
\end{document}